\documentclass[pdflatex,sn-standardnature]{sn-jnl}

\usepackage{ulem}

\usepackage{comment}
\long\def\comment#1{}
\newcommand{\review}{}

\newcommand{\Msun}{M$_{\odot}$} 
 
\newcommand{\Lsun}{L$_{\odot}$}

\newcommand{\mic}{{$\mu$m}}
\newcommand{\kms}{km~s$^{-1}$}

\newcommand{\nii}{N~{\sc ii}}
\newcommand{\hh}{H$_2$}

\newcommand{\siii}{S~{\sc iii}}

\newcommand{\neii}{Ne~{\sc ii}}

\newcommand{\araa}{Annu. Rev. Astron. Astrophys.}   
\newcommand{\aj}{Astron. J.}   
\newcommand{\apj}{Astrophys. J.}   
\newcommand{\apjl}{Astrophys. J. Lett.}   
\newcommand{\apjs}{Astrophys. J. Suppl. Ser.}   
\newcommand{\aap}{Astron. Astrophys.}   
\newcommand{\aapr}{Astron. Astrophys. Rev.}   
\newcommand{\mnras}{Mon. Not. R. Astron. Soc.}   
\newcommand{\nat}{Nature} 
\usepackage{xcolor}

\jyear{2021}%

\theoremstyle{thmstyleone}%
\theoremstyle{thmstyletwo}%

\theoremstyle{thmstylethree}%

\begin{document}

\title[JWST PN]%
{The messy death of a multiple star system and the resulting planetary nebula as observed by JWST}




\author*[1,2]{\fnm{Orsola} \sur{De Marco}}\email{orsola.demarco@mq.edu.au}

\author[3,4]{\fnm{Muhammad} \sur{Akashi}}

\author[5]{Stavros Akras}

\author[6]{Javier Alcolea} 

\author[7]{Isabel Aleman} 

\author[8]{Philippe Amram} 

\author[9]{Bruce Balick} 

\author[10]{Elvire De Beck} 

\author[11,12]{Eric G. Blackman}

\author[13]{Henri M. J. Boffin}

\author[5]{Panos Boumis} 

\author[14]{Jesse Bublitz} 

\author[15]{Beatrice Bucciarelli}

\author[6]{Valentin Bujarrabal} 

\author[16,17,18]{Jan Cami}

\author[19]{Nicholas Chornay}

\author[20]{You-Hua Chu}

\author[21,22]{Romano L.M. Corradi} 

\author[23]{Adam Frank}

\author[24]{Guillermo Garc\'{i}a-Segura}

\author[22,25]{D. A. Garc\'{i}a-Hern\'{a}ndez}

\author[22,25]{Jorge Garc\'{i}a-Rojas}

\author[22,25]{Veronica G\'{o}mez-Llanos}

\author[26]{Denise R. Gon\c{c}alves}

\author[27]{Mart\'{i}n A. Guerrero} 

\author[22,25]{David Jones} 

\author[28,29]{Amanda I. Karakas} 

\author[30,31]{Joel H. Kastner} 

\author[32]{Sun Kwok} 

\author[33,34]{Foteini Lykou} 

\author[22,25,35]{Arturo Manchado}

\author[36]{Mikako Matsuura}

\author[37,38]{Iain McDonald}

\author[39]{Ana Monreal-Ibero}

\author[7]{Hektor Monteiro}

\author[31]{Paula Moraga Baez}  

\author[24]{Christophe Morisset}

\author[40]{Brent Miszalski} 

\author[41,42,43,44]{Shazrene S. Mohamed}

\author[45]{Rodolfo Montez Jr.} 

\author[46,47]{Jason Nordhaus}

\author[48]{Claudia Mendes de Oliveira}

\author[28,29]{Zara Osborn} 

\author[49]{Masaaki Otsuka}

\author[50,51]{Quentin A. Parker}

\author[16,17,18]{Els Peeters}

\author[52]{Bruno C. Quint}

\author[53]{Guillermo Quintana-Lacaci}

\author[54]{Matt Redman}

\author[55]{Ashley J. Ruiter}

\author[24]{Laurence Sabin}

\author[56]{Carmen S\'{a}nchez Contreras}

\author[6]{Miguel Santander-Garc\'{i}a}

\author[55]{Ivo Seitenzahl}

\author[57]{Raghvendra Sahai}

\author[3]{Noam Soker}

\author[58]{Angela K. Speck}

\author[59]{Letizia Stanghellini}

\author[60]{Wolfgang Steffen}

\author[61]{Jes\'{u}s A. Toal\'{a}}

\author[62]{Toshiya Ueta} 

\author[63]{Griet Van de Steene}

\author[56]{Eva Villaver}

\author[64]{Paolo Ventura}

\author[65]{Wouter Vlemmings}

\author[13]{Jeremy R. Walsh}

\author[36]{Roger Wesson}

\author[66]{Hans van Winckel}

\author[38]{Albert A. Zijlstra}


\affil*[1]{School of Mathematical and Physical Sciences, Macquarie University, Sydney, NSW 2109, Australia} 

\affil*[2]{Astronomy, Astrophysics and Astrophotonics Research Centre, Macquarie University, Sydney, NSW 2109, Australia} 

\affil[3]{Department of Physics, Technion, Haifa, 3200003, Israel} 

\affil[4]{Kinneret College on the Sea of Galilee, Samakh 15132, Israel} 

\affil[5]{Institute for Astronomy, Astrophysics, Space Applications and Remote Sensing, National Observatory of Athens, GR 15236 Penteli, Greece} 

\affil[6]{Observatorio Astron\'{o}mico Nacional (OAN/IGN), Alfonso XII, 3, 28014 Madrid, Spain} 

\affil[7]{Instituto de F\'{i}sica e Qu\'{i}mica, Universidade Federal de Itajub\'{a}, Av. BPS 1303, Pinheirinho, Itajub\'{a} 37500-903, Brazil} 

\affil[8]{Aix-Marseille Univ., CNRS, CNES, LAM (Laboratoire d’Astrophysique de Marseille), Marseille, France} 

\affil[9]{Astronomy Department, University of Washington, Seattle, WA 98105-1580, USA} 

\affil[10]{Department of Space, Earth and Environment, Chalmers University of Technology, S-41296 Gothenburg, Sweden} 

\affil[11]{Department of Physics and Astronomy, University of Rochester, Rochester, NY 14627, USA} 

\affil[12]{Laboratory for Laser Energetics, University of Rochester, Rochester NY, 14623, USA} 

\affil[13]{European Southern Observatory, Karl-Schwarzschild Strasse 2, D-85748 Garching, Germany} 

\affil[14]{Green Bank Observatory, 155 Observatory Road, PO Box 2, Green Bank, WV 24944, USA} 

\affil[15]{INAF - Osservatorio Astrofisico di Torino, Via Osservatorio 20, 10023, Pino Torinese, Italy} 

\affil[16]{Department of Physics \& Astronomy, University of Western Ontario, London, ON, N6A 3K7, Canada}

\affil[17]{Institute for Earth and Space Exploration, University of Western Ontario, London, ON, N6A 3K7, Canada}

\affil[18]{SETI Institute, 399 Bernardo Avenue, Suite 200, Mountain View, CA 94043, USA}

\affil[19]{Institute of Astronomy, University of Cambridge, Madingley Road, Cambridge CB3 0HA, UK} 

\affil[20]{Institute of Astronomy and Astrophysics, Academia Sinica (ASIAA), No. 1, Section 4, Roosevelt Road, Taipei 10617, Taiwan}

\affil[21]{GRANTECAN, Cuesta de San Jos\'{e} s/n, E-38712, Bre\~{n}a Baja, La Palma, Spain} 

\affil[22]{Instituto de Astrof\'isica de Canarias, E-38205 La Laguna, Tenerife, Spain} 

\affil[23]{Department of Physics and Astronomy, University of Rochester, Rochester, NY 14627, USA} 

\affil[24]{Instituto de Astronom\'{i}a, Universidad Nacional Aut\'{o}noma de M\'{e}xico, Km. 107 Carr. Tijuana-Ensenada, 22860, Ensenada, B.~C., Mexico} 

\affil[25]{Departamento de Astrof\'{i}sica, Universidad de La Laguna, E-38206 La Laguna, Tenerife, Spain} 

\affil[26]{Observat\'{o}rio do Valongo, Universidade Federal do Rio de Janeiro, Ladeira Pedro Antonio 43, Rio de Janeiro 20080-090, Brazil}

\affil[27]{Instituto de Astrof\'{i}sica de Andaluc\'{i}a, IAA-CSIC, Glorieta de la Astronom\'{i}a, s/n, E-18008, Granada, Spain} 

\affil[28]{School of Physics \& Astronomy, Monash University, Clayton VIC 3800, Australia} 

\affil[29]{ARC Centre of Excellence for All Sky Astrophysics in 3 Dimensions (ASTRO 3D)} 

\affil[30]{Center for Imaging Science, Rochester Institute of Technology, Rochester, NY 14623, USA} 

\affil[31]{School of Physics and Astronomy and Laboratory for Multiwavelength Astrophysics, Rochester Institute of Technology, USA} 

\affil[32]{Department of Earth, Ocean, and Atmospheric Sciences, University of British Columbia, Vancouver, Canada} 

\affil[33]{Konkoly Observatory, Research Centre for Astronomy and Earth Sciences, E\"otv\"os Lor\'and Research Network (ELKH), Konkoly-Thege Mikl\'os \'ut 15-17, 1121 Budapest, Hungary} 

\affil[34]{CSFK, MTA Centre of Excellence, Konkoly-Thege Mikl\'os \'ut 15-17, 1121 Budapest, Hungary} 

\affil[35]{Consejo Superior de Investigaciones Cient\'{\i}ficas, Spain} 

\affil[36]{School of Physics and Astronomy, Cardiff University, The Parade, Cardiff CF24 3AA, UK} 

\affil[37]{Department of Physical Sciences, The Open University, Walton Hall, Milton Keynes, MK7 6AA, UK} 

\affil[38]{Jodrell Bank Centre for Astrophysics, Department of Physics and Astronomy, The University of Manchester, Oxford Road M13 9PL Manchester, UK} 

\affil[39]{Leiden Observatory, Leiden University, Niels Bohrweg 2, NL 2333 CA Leiden, The Netherlands} 

\affil[40]{Australian Astronomical Optics, Faculty of Science and Engineering, Macquarie University, North Ryde, NSW 2113, Australia} 

\affil[41]{Department of Physics, University of Miami, Coral Gables, FL 33124, USA}

\affil[42]{South African Astronomical Observatory, P.O. Box 9, 7935 Observatory, South Africa}

\affil[43]{Astronomy Department, University of Cape Town, 7701 Rondebosch, South Africa}

\affil[44]{NITheCS National Institute for Theoretical and Computational Sciences, South Africa}

\affil[45]{Center for Astrophysics, Harvard \& Smithsonian, 60 Garden Street, Cambridge, MA 02138, USA} 

\affil[46]{Center for Computational Relativity and Gravitation, Rochester Institute of Technology, Rochester, NY 14623, USA} 

\affil[47]{National Technical Institute for the Deaf, Rochester Institute of Technology, Rochester, NY 14623, USA} 

\affil[48]{Departamento de Astronomia, Instituto de Astronomia, Geof\'isica e Ci\^encias Atmosf\'ericas da USP, Cidade Universit\'aria, 05508-900, S\~ao Paulo, SP, Brazil} 

\affil[49]{Okayama Observatory, Kyoto University, Honjo, Kamogata, Asakuchi, Okayama, 719-0232, Japan}

\affil[50]{Department of Physics, CYM Physics Building, The University of Hong Kong, Pokfulam, Hong Kong SAR, PRC} 

\affil[51]{Laboratory for Space Research, Cyberport 4, Cyberport, Hong Kong SAR, PRC} 

\affil[52]{Rubin Observatory Project Office, 950 N. Cherry Ave., Tucson, AZ 85719, USA} 

\affil[53]{Dept. of Molecular Astrophysics. IFF-CSIC. C/ Serrano 123, E-28006, Madrid, Spain} 

\affil[54]{Centre for Astronomy, School of Physics, National University of Ireland Galway, Galway H91 CF50, Ireland}

\affil[55]{University of New South Wales, Australian Defence Force Academy, Canberra, Australian Capital Territory, Australia}

\affil[56]{Centro de Astrobiolog\'{i}a (CAB), CSIC-INTA, Camino Bajo del Castillo s/n, ESAC campus, 28692, Villanueva de la Ca\~{n}ada, Madrid, Spain} 

\affil[57]{Jet Propulsion Laboratory, California Institute of Technology, CA 91109, Pasadena, USA} 

\affil[58]{University of Texas at San Antonio, Department of Physics and Astronomy, Applied Engineering and Technology Building, One UTSA Circle, San Antonio, TX 78249, United States}

\affil[59]{NSF's NOIRLab, 950 N. Cherry Ave., Tucson, AZ 85719, USA} 

\affil[60]{ilumbra, AstroPhysical MediaStudio, Hautzenbergstrasse 1, 67661 Kaiserslautern, Germany} 

\affil[61]{Instituto de Radioastronom\'{i}a y Astrof\'{i}sica, UNAM, Antigua Carretera a P\'{a}tzcuaro 8701, Ex-Hda. San Jos\'{e} de la Huerta, Morelia 58089, Mich., Mexico} 

\affil[62]{Department of Physics and Astronomy, University of Denver, 2112 E Wesley Ave., Denver, CO 80208, USA} 

\affil[63]{Royal Observatory of Belgium, Astronomy and Astrophysics, Ringlaan 3, 1180 Brussels, Belgium} 

\affil[64]{INAF -- Osservatorio Astronomico di Roma, Via Frascati 33, I-00040, Monte Porzio Catone (RM), Italy}

\affil[65]{Onsala Space Observatory, Department of Space, Earth and Environment, Chalmers University of Technology, Onsala, Sweden} 

\affil[66]{Institute of Astronomy, KULeuven, Celestijnenlaan 200D, B-3001 Leuven, Belgium} 

\abstract{
Planetary nebulae (PNe), the ejected envelopes of red giant stars, provide us with a history of the last, mass-losing phases of 90\%\ of stars initially more massive than the Sun. Here, we analyse James Webb Space Telescope (JWST) Early Release Observation (ERO) images of the\comment{\sout{nearby}} PN NGC~3132. A\comment{\sout{highly}} structured, extended \hh\ halo surrounding an ionised\comment{\sout{PN}} central bubble is imprinted with spiral structures, likely shaped by a low-mass companion orbiting the central star at $\sim$40--60~AU. The images also reveal a mid-IR excess at the central star interpreted as a dusty disk, indicative of an interaction with another, closer companion. Including the previously known, A-type visual companion, the progenitor of the NGC~3132 PN must have been at least a stellar quartet.\comment{\sout{Details of the flocculent halo}} \review{The JWST images} allow us to generate a model of the illumination, ionisation and hydrodynamics of the molecular halo, demonstrating the power of JWST to investigate complex stellar outflows. Further, new measurements of the A-type visual companion allow us to derive  the \comment{\sout{most precise}} value for the mass of the progenitor of a central star to date \review{with excellent precision}: $2.86\pm0.06\,\rm M_\odot$ \review{}.\comment{\sout{This value imposes a much needed constraint for the minimum mass for the operation of hot-bottom burning. }}
These\comment{\sout{ERO}} results 
\comment{\sout{for NGC~3132}} serve as pathfinders for future JWST observations of PNe providing unique insight into fundamental astrophysical processes including colliding winds,\comment{\sout{plasma instabilities,}} and binary star interactions, with implications for supernovae and gravitational wave systems.}




\maketitle

\section*{Main}

\subsection*{Introduction}
\label{sec:introduction}

Planetary nebulae (PNe) are the ejected envelopes of intermediate-mass ($\sim$1--8~\Msun) stars that have recently terminated their asymptotic giant branch (AGB) stage of evolution. Moving outwards from the hot pre-white dwarf star ($T\sim 10^5$\,K) that is the progeny of the AGB star, the structure of a canonical quasi-spherical PN consists of a hot, sparse, wind-heated bubble ($T\sim 10^7$K) surrounded by a dense shell of displaced, ionised AGB gas ($T\sim 10^4$\,K), which in turn may still be surrounded by ``pristine,'' cold  ($T\sim 10^2$\,K), molecule- and dust-rich AGB ejecta. On the other hand, if the progenitor star interacted with a companion(s) during its post-main sequence evolution, we would expect departures from spherical symmetry, perhaps including spiral structures and arcs \citep[e.g.,][]{Mastrodemos1999,Mohamed2012,Maercker2012}, the presence of a dense, molecule-rich torus \citep[e.g.,][]{Santander-Garcia2017}, one or more pairs of polar lobes formed by fast, collimated outflows and jets \citep[e.g.,][]{Sahai1998, Sahai11}, and/or a dusty, circumbinary disk \cite{VanWinckel2003}. The type of interaction depends on the orbital radius, and ranges from common envelope evolution for close binaries \cite{Ivanova2013}, to accretion disks and gravitational focussing of the wind for wider systems \cite{Mastrodemos1998,Mohamed2007,deValBorro2009}, to displacement of the central star from the geometric centre of the nebula for the widest systems \cite{Soker1999b}.

The first Hubble Space Telescope (HST) images of PNe revealed a breathtaking new world of details and far more complex structures than had been gleaned from ground-based images \citep[e.g.,][]{Balick1998,Sahai98}. The superb spatial resolution of HST, combined with high-resolution, kinematic mapping, enabled the construction of detailed 3D, morpho-kinematic models, which, together with hydrodynamic models \citep[e.g.,][]{Sabbadin2006,Steffen2006}, started to connect our understanding of the evolution of the structures and kinematics of PNe with their possible binary star origins \citep[e.g.,][]{Balick2002,DeMarco2009,Jones2017}.

The James Webb Space Telescope (JWST), with its superb sensitivity and high spatial resolution from near- to mid-IR, is now poised to enable a leap of similar magnitude in our understanding of PNe. This journey began when JWST released near-IR and mid-IR images of just one PN, NGC~3132, as part of its ERO program.  NGC~3132 is a nearby ($D \sim 750$ pc), molecule-rich \cite{Sahai1990,Kastner1996}, ring-like PN, long known to harbour a visual binary comprising the central (progenitor) star and an A star companion. In this paper we show that the JWST ERO images  contain multiple, new lines of evidence that NGC~3132 is the recent product of a hierarchical multiple progenitor stellar system, which has experienced both indirect and direct interactions involving one or more components. Such binary interactions have taken on new importance in the era of gravitational wave detectors (LIGO \cite{Abramovici1992}, LISA \cite{AmeroSeoane2017}) and ambitious transient surveys \cite{Ivezic2008}. Indeed, PNe like NGC~3132 offer unique insight into the formation pathways of the close, single and double degenerate binaries that are eventual gravitational wave sources and (perhaps) type Ia supernova progenitors \citep{SantanderGarcia2015,Chiotellis2020, Cikota2017}.

\subsection*{Results}
\label{sec:results}

\comment{\sout{As part of its ERO program \citep{JWST_ERO}, JWST obtained ten images of NGC~3132 (six NIRCam images and four MIRI images) through broadband and narrow-band filters at wavelengths from 0.9 to 18 $\mu$m. The full resulting JWST image suite, along with basic information about NIRCam and MIRI image formats and filters, is presented in Specification of JWST NIRCam and MIRI imaging and Supplementary Figure~1. The NIRCam and MIRI filters isolate specific ionic lines and rovibrational transitions of H$_2$; they are relatively unaffected by foreground and especially intranebular extinction, and cover the thermal IR. The subarcsecond-resolution JWST infrared images offer an unimpeded view of the stratified excitation structure of NGC~3132, readily distinguishing between the nebula's ionised and molecular gas components, and revealing for the first time a track sculpted by an unseen companion (Figure~\ref{fig:HIIvsH2}). Furthermore, the JWST MIRI images reveal the presence of warm dust at the location of the central star, easily reconcilable with a dusty disk, likely the result of a close binary interaction (Figure~\ref{fig:CSPNimages_IRexcess}). }}

\subsubsection*{A flocculent molecular halo surrounding an ionised bubble}\label{sec3}

Figure~\ref{fig:HIIvsH2} displays colour overlays of NIRCam and MIRI images of NGC~3132 that highlight JWST's clean separation of the PN's ionised (H~{\sc ii}) and molecular (H$_2$) regions. \review{The full resulting JWST image suite, along with basic information, is presented in Specification of JWST NIRCam and MIRI imaging and Supplementary Figure~1. The images reveal, for the first time, the extent and detailed structure of the halo of molecular gas that lies exterior to the nebula's central, ionised cavity and its bright and thin, peripheral elliptical ring (cf. \cite{Hora2004}). This molecular halo is well detected in rovibrational H$_2$ emission at 2.12 \mic\ (1--0 S(1)), 4.7 \mic\ (0--0 S(9)), and 7.7 \mic\ (0--0 S(5)) out to 60~arcsec ($\sim$0.22 pc at the adopted distance of 754~pc, see Properties and distance of NGC~3132) from the central star. Spatially organised structures --- arcs and patterns of spikes emanating radially outward 
\comment{\sout{from the central star}} --- are observed in the halo H$_2$ emission on medium to large scales, while molecular arcs, loops, and knots are detected on size scales from $\sim$500 AU down to the limiting ($\sim$75 AU ) resolution of the images. The typical thickness of the bright H$_2$ rings that surround the nebular core is $\sim$\comment{\sout{0.15}} $1-2$~arcsec \review{($\sim$750-1500~AU), measured at 2.1, 4.7 and 7.7~\mic}. 

Figure~\ref{fig:HIIvsH2} conclusively demonstrates that the molecular gas is much clumpier than the ionised gas component of NGC~3132 (see also Supplementary  Figure~3). In hydrogen recombination lines and [S~{\sc iii}] emission (Figure~\ref{fig:HIIvsH2}, top-left), the nebula's central ionised cavity (within $\sim$25 arcsec of the central star) appears as a relatively smooth elliptical region that is bounded by a single, sharped-edged ring; whereas in \hh\ (Figure~\ref{fig:HIIvsH2}, bottom-left), this same central region appears as a far more complex system of clumpy filaments.
The regions in and around this bright, inner H$_2$ ring system contain as many as 20 dense clumps (knots) per square arcsec, implying the total number of H$_2$ knots in this region exceeds $10^4$. The H$_2$ knots in the outer (halo) region are less distinct and further apart. 

The \review{presence of} radially-directed spike features in the \hh\ halo\comment{\sout{all precisely point to the ionising star. This}} indicates that direct irradiation by UV photons, leaking through less dense gas between the inner ring system's H$_2$ knots, are most likely responsible for the excitation of the IR H$_2$ lines in the extended halo, although shock excitation cannot be completely ruled out (see \cite{Fang18} and references therein). The relative lack of H$_2$ halo emission to the East-Northeast and West-Southwest of the central star then indicates a general lack of central star UV illumination, as opposed to lack of halo molecular mass in those directions (see Discussion). Measurements of the extinction of background nebulosity through representative knots suggests typical knot densities of $\sim10^6$ cm$^{-3}$ and masses of $\sim 10^{-5}$\,M$_\odot$ (see Densities, masses and excitation of the \hh\ knots), suggesting a total H$_2$ mass of $\sim$0.1~M$_\odot$ in the central ring region. 

The system of (broken) concentric arcs revealed in the \hh\ halo by the JWST images is similar to those observed in the extended, dusty envelopes of many AGB stars, proto-PNe and PNe (e.g., \cite{Maercker2012,RamosLarios2016,Guerreroetal2020}).  A widely accepted scenario to explain the formation of such arc systems is the modulation of an AGB wind by a stellar or substellar companion, creating 3D spiral-like patterns along the orbital plane \citep[see][and references therein]{Mastrodemos1999,Kimetal2019,Maes2021,Aydi2022,Decin20}. The average angular distance between the arc structures, 2~arcsec, implies an orbital period of 290-480~years and an orbital separation of 40-60~AU between the central star and the companion that shapes the mass loss. Here, we have assumed a companion mass of 0.2~\Msun, the highest mass main-sequence star that could hide in the present-day central star's glare yet still form a visible arc system (other parameters are an expansion velocity in the range 15-25~\kms\ \citep{Guerreroetal2020} and an assumed late-AGB central star mass of \review{$\sim$}0.8~\Msun; \review{likely still 0.1-0.2~\Msun\ larger than the post-AGB mass}). The bright A2~V visual companion seen at $\sim$1300\,AU projected separation from the central star cannot be responsible, suggesting (at least) a triple system in a stable configuration.

\subsubsection*{The dusty central star}
\label{sec:results-dusty-central-star}

In the MIRI images obtained at wavelengths longer than 10~\mic, the faint central star appears as bright or brighter than its A2~V main sequence visual companion \citep{Mendez1978}; see Figure~\ref{fig:CSPNimages_IRexcess}. This infrared excess was undetectable in the mid-infrared at lower spatial resolution (e.g., in WISE images \cite{Wright2010}) because of\comment{\sout{the proximity of the A-type companion and}} the surrounding bright nebulosity. The JWST-discovered IR excess indicates that a considerable amount of warm dust is present around the ultra-hot ($\sim$110 kK) PN central star. The thermal infrared source appears marginally extended in the 11.3 and 12.8~\mic\ MIRI images with an apparent size of $\sim$300~AU (FWHM) at 12.8~\mic\ (see PSF measurements of the central star).

The bottom panel of Figure~\ref{fig:CSPNimages_IRexcess} displays the central star's near-IR to mid-IR spectral energy distribution fitted by a combination of  a hot stellar photosphere represented by a blackbody curve and\comment{\sout{and dusty disk}} \review{two curves to fit the infrared data points.} The two curves are generated with a model that follows closely that of \cite{Su2007} for the Helix nebula. \review{A number of 100~\mic\ grains are taken as blackbody spheres with temperatures set by absorption and re-emittance of the stellar luminosity (200~\Lsun; a correction factor is then applied to simulate a grain size distribution between 60 and 1000~\mic, as done by \cite{Su2007}). The temperature varies as $d^{0.5}$, where $d$ is the distance to the star. The surface density of the disk is taken as constant. The resulting blackbody radiation is calculated at each radius, and the emission is summed over all radii. A better model  will require radiative transfer, actual dust emissivities, a range of grains sizes, and for the silicate feature, the inclination of the disk. This will be explored in a future paper.  }

The best-fit model disk has an inner radius of 55\,AU and outer radius 140\,AU, and a dust mass of $3\times10^{26}$~g or $2\times10^{-7}$~\Msun\ (approximately\comment{\sout{one}} \review{0.05} Earth masses). The dust temperature range (inner to outer radius) is  130 to 80\,K. The outer radius of 140~AU, though poorly constrained, is consistent with the deconvolved half-width of the marginally extended mid-IR source. These dimensions resemble those inferred for the disk orbiting the central star of the Helix (35--150\,AU; \cite{Su2007}), but the dust mass is somewhat\comment{\sout{larger}} \review{smaller} (cfr.\comment{\sout{$3\times10^{-7}$~\Msun}} \review{0.13 earth masses}). The outer radius could be slightly larger, if the 18~\mic\ flux is underestimated because of detector saturation. An additional inner, hotter disk --- with radius between 3 and 8\,AU, a temperature between 550 to 335~K (inside to outside) and a very small mass of  \comment{\sout{$4\times10^{23}$~g}} $2\times10^{22}$~g (approximately \review{0.02 times the} mass of Ceres) --- is needed to fit the 3.5 and 7~\mic\ fluxes.  \review{While this model does not constrain the geometry of the distribution to be that of a disk, the reasoning behind a disk structure is based on a physical reasoning whereby only a rotating Keplerian disk can be shown to be stable and relatively long-lived, while other structures, such as shells, are easily shown to be unstable \cite{Clayton2014}.}

The A2~V companion is slightly evolved \citep{Mendez1978} and has a mass of $M_{\rm A2V}=2.40\pm0.15\,\rm M_\odot$, using the PARSEC isochrones. Its visual companion, the PN central star, must have descended from a more massive star, as it has evolved faster. Extrapolating the same PARSEC isochrone gives an initial main sequence mass for the central star of $M_{\rm i}=2.86\pm 0.06\,\rm M_\odot$. This is potentially the most precise initial mass for any PN central star or white dwarf yet determined.
\comment{\sout{with the caveat that the model isochrones may themselves be subject to additional uncertainties. }}
\review{We estimate the error to be 0.16~\Msun\ if we add systematic effects between different isochrone models (see Central star system's masses).}

\comment{\sout{Furthermore, 
this mass is consistent with the one derived by the nebular abundance and the stellar evolutionary yields (see Methods, Section~\ref{sec:methods-central-star-mass}).
the measured N/O ratio of 0.4 indicates that hot bottom burning has not taken place in this star, imposing a rare and much needed constraint on the lower mass limit for the occurrence of hot-bottom burning (current models place that lower limit at $\sim$3~\Msun\ and $\sim$4~\Msun\ for the ATON and MONASH models, respectively showing in \citep{Ventura2018}.The gas phase C/O measurements of the PN are not reliable, but there is evidence that the dust in the PN is oxygen-rich rather than carbon-rich \citep{Delgado-Inglada2014}, an indication that the C/O ratio of the gas phase is below unity. The same models \citep{Ventura2018} predict that C/O ratios $<$1 are achieved via hot bottom burning (HBB) for masses of $\sim$3~\Msun\ (ATON models) or $>$4.75~\Msun\ (MONASH models).  Our mass measurement of 2.86~\Msun, therefore, would be more in line with the ATON models, if we allow for some uncertainty.}} 

The current (near-final) mass of a PN central star descended from such a $\sim$2.9 \Msun\ progenitor is predicted to be\comment{\sout{$M_f \sim 0.71\pm0.05$\,M$_\odot$ based on initial-final mass relations \cite{Bertolami2016}}} \review{$M_f \sim 0.66\pm0.05$\,M$_\odot$ based on initial-final mass relations \cite{Ventura2018}}, albeit with larger systematic uncertainties that are dependent on details of the mass loss process adopted by the models. It is noteworthy that photoionisation models of the nebula require a cooler, dimmer and overall less massive central star \review{(0.58$\pm$0.03~\Msun)} than what we have found.\comment{\sout{; however, after accounting for the extinction due to the dusty disk detected by JWST, the central star mass is better reconciled with what is inferred from the presence of its A type companion (see Methods, Section~\ref{sec:methods-photoionisation-modelling}).} }

\review{We find that we can reconcile the mass of the star today and that of the  photoionisation model, while also matching the nebular abundances and the nebular age, if we assume that the AGB evolution of a  2.86~\Msun\ star, was interrupted by a binary interaction that ejected the envelope. We conjecture that the AGB evolution was interrupted at a core mass of 0.61~\Msun, because for larger values, the C/O ratio of the stellar envelope gas would increase above unity (counter to the observation of crystalline silicate grains). At larger masses the N/O ratio would also increase above the observed value of 0.42.}

\subsection*{Discussion}
\label{sec:discussion}

The first striking discovery of JWST is the presence of the dusty disk around the ultra-hot central star. This indicates that JWST can accurately detect dusty disks lighter than Ceres, as far as $\sim$700~pc away. For our PN, the presence of such a disk orbiting the PN central star favours a close binary interaction, where the companion either merged with the primary star, or is still in orbit but is undetected (mass $<$ 0.2~\Msun; \review{based on an unresolved or barely resolved, equal-brightness companion}); in either case, the companion has donated a substantial fraction of its angular momentum to the gas \cite{Huang1963,Soberman1997}. Observationally, such disks around PN central stars, though rare, appear to be by and large associated with known or strongly suspected binarity \cite{Clayton2014} and may be related to circumbinary disks detected around other classes of post-AGB binary stars 
\cite{vanWinckel2009}.

An interacting binary scenario is reinforced by the shape of the ionised cavity, which represents the inner, most
recent mass-loss phase, when the already hot central star emitted a fast, tenuous wind. Pairing the JWST images with spatially resolved spectroscopy we constructed a 3D visualisation of this cavity (see Morpho-kinematic modelling in Supplementary Material).
In Figure~\ref{fig:de-project_shape_views} we show that this inner cavity is inferred to be an expanding prolate ellipsoid\comment{\sout{viewed almost pole-on}} \review{with its long axis tilted at approximately 30~$\deg$ to the line of sight}.  Its surface is not smooth and presents instead a number of protuberances, most of which can be paired via axes passing through, or very near the central star. \comment{\sout{In our 3D reconstruction, NGC~3132's bright inner ring system is decomposed into a belt structure at approximately zero Doppler shift, that is misaligned with the waist of the ellipsoid, as well as an additional single filament (an uneven ring) wrapping around the ellipsoid, but that is misaligned with both the ellipsoid waist and the belt.}} Prolate cavities such as these, with misaligned structures, are common in PN and are likely sculpted by jets from interacting binaries in the earlier, pre-PN phase of the nebula \cite{Sahai2000}, with additional details added during the interaction between the AGB wind and post-AGB fast wind and via the process of PN ionisation. 

The numerous protuberances clearly evident in the 3D reconstruction could arise from ionised gas breaking out of the inner cavity through an uneven outer shell. The apparent pairing between these protuberances may argue instead for the presence of intermittent and toppling jets \citep{AkashiSoker2021}.  To generate jets over such a wide range of axes, an interacting binary is not enough, and one would have to conjecture that the central star is or was a member of not just a close binary, but of an interacting {\it triple} system \citep{BearSoker2017}. Recent studies of interactions in triple systems \citep{Hamers2022,Glanz2021} also argue for the possibility of interactions yielding complex ejecta.

Outside the ionised ellipsoid, one encounters material ejected earlier in the star's history. The AGB mass loss, at rates of up to $\sim$10$^{-5}$ \Msun yr$^{-1}$ and speeds of $\sim$10~km~s$^{-1}$ over a $\sim$10$^5$~yr timescale \citep{HofnerOlofsson2018}, generates an enormous, expanding envelope of molecular gas and dust. The \hh\ halo imaged by JWST constitutes the most recently ejected (inner) region of this AGB envelope. 
The\comment{\sout{straight}} spikes observed in the halo (Figure~\ref{fig:HIIvsH2}, right panel) show that the inner cavity is very porous, though less so near the minor axis where the cavity edges are brightest, densest, and least fractured.

The JWST images motivated 2D hydrodynamic simulations to replicate these flocculent structures.   In Figure~\ref{fig:guille-fragmentation2} we see two time snapshots towards the end of a simulation where an inner, faster wind from the heating central star and its ionising radiation, plough into the dense AGB (halo) material (see Methods, Section~\ref{sec:methods-hydro-modelling}). The fragmentation that happens at the interface of the swept-up material also creates the variable opacity needed \review{to shield some of the wind material from ionising radiation, which then quickly recombines and allows the formation of molecules. Non ionising radiation leaks more readily because the opacity above 913~\AA\ is lower. These photons produce florescence of \hh.}
\comment{\sout{to allow the central star radiation to leak out along discrete directions, causing illumination spikes.} }

In Figure~\ref{fig:guille-fragmentation2} we see two time snapshots towards the end of the simulation. In the first panel we see a set of \review{approximately} radial spikes, but 200 years later those straight and thin spikes  evolve to thicker and sometimes curved ones. In the right column of Figure~\ref{fig:guille-fragmentation2} two different parts of the nebula exhibit thinner and straighter spikes (top-right panel) or thicker, bent ones (bottom-right panel). Although the entire nebula was ejected and ionized over a short time interval, there can be a delay in the evolution of a given spike in a specific part of the nebula,  related to the local opacity in the swept-up shell. 
Figure~\ref{fig:guille-fragmentation2} suggests that differences of only $\sim$ 200 years in the timescales of mass ejection and/or the progress of illumination along specific directions can explain the marked differences observed in the flocculent structure around the nebula.

The successful modelling of illumination percolating unevenly into the molecular halo (Figure~\ref{fig:guille-fragmentation2})  motivated a further geometric model of the halo, presented in Figure~\ref{fig:h2-and-comparison}. This Figure compares the extended H$_2$ structures as imaged by JWST with a  model consisting of two thick, concentric, unbroken but clumpy, shells of material that are illuminated by the central star through a porous ellipsoid representing the boundary of the ionised cavity, with reduced opacity in the polar regions. 
As a result of the uneven illumination the distribution of \hh\ material appears fragmented and is generally brighter toward the polar regions (and suppressed along the equatorial plane) of the central ellipsoidal, ionised region. The distribution seen in the JWST \hh\ images could be reproduced more closely by altering the opacity of the inner ellipsoid. Fly-though movies of the 3D reconstructions of both the inner ellipsoid (\href{www.ilumbra.com/public/science/ilumbra_NGC3132_NII_reconstruction.mp4}{Figures~\ref{fig:de-project_shape_views}}) and the outer \hh\ halo (\href{https://www.ilumbra.com/public/science/H2_Shape_scattering.mp4}{Figure~\ref{fig:h2-and-comparison}}) can be found following the links.

The arches in the JWST images, are not smeared as is typical of those seen in projection \citep[e.g.,][]{Balick2012}, but are instead sharp. This possibly indicates that these arches are on or near the plane of the sky, indicating that the orbit of the companion at $\sim$40-60~AU is closely aligned to the waist of the inner ellipsoid. This companion cannot partake in the formation of the disk around the central star, though it may play a secondary role in the shaping of other PN structures. It is also unlikely to have launched strong jets because at such distance the accretion rate would be very low. As such, this would be an additional companion to the inner binary (or triple), making it a tertiary (or quaternary) companion. 

The visual A-type companion would then be a fourth (fifth) member of the group, an almost complete bystander from the point of view of interaction and shaping, but critically important for this study: Its well measured mass, and slight evolved status, constrained the initial mass of the central star: (2.86$\pm$0.06)~\Msun.\comment{\sout{This and the N/O ratio of the PN, impose a tight constraint on the lower mass for the occurrence of hot bottom burning, a hotly debated topic in stellar evolution.}}

To reconstruct the events that lead to the demise of the progenitor of NGC3132, the PN acts like a murder scene. The A-type companion, could not have partaken to the interaction that unravelled the AGB star, but was (and is) certainly present. A second companion at 40-60AU left an indelible trail of its presence in the form of arcs, but was not close enough to generate the dusty disk, nor shape the ionised cavity, implying that there must have been at least another accomplice. This points the finger at a close-by companion, that is either avoiding detection, or has perished in the interaction  (merged). If the numerous protuberances seen in the ionised cavity come in pairs, then tumbling jet axes would be needed and this would point the finger to the presence of a {\it second}, close companion \citep{Hamers2022,Glanz2021}, which would make the system a quintet. Even ignoring the putative second, close companion, we can state with good degree of certainty that the system is at least a quartet. Systems of four or five stars orbiting within a few $\times$1000 AU are not impossibly rare for primary stars in the progenitor mass range of interest here \citep[e.g., HD 104237;][]{Feigelson2003}; indeed, present estimates indicate that 50\%\ or more of stars of 2-3~\Msun\ are in multiple systems, and of order 2\% of A-type stars have four companions \citep{Duchene2013}. 

JWST is at the starting gate of its promise as an astrophysical pathfinder. With complementary radio, interferometric and time resolved observations, it can find the temporal signatures of active convective mass ejection from the surfaces of AGB stars and the subsequent gravitational influence of companion stars in dynamically- and thermally-complex outflows.  
Thus JWST offers the potential to intimately connect the histories of PNe and the role of close stellar companions to studies of chemical evolution, nebular shaping and binary interactions for the next century.

\section*{Methods}
\label{sec:methods}

\subsection*{Properties and distance of NGC~3132}
\label{sec:methods-distance}

The inner, ionised cavity of NGC~3132 is elliptical in shape, with a major axis of $\sim$40 arcsec (0.15 pc) and an electron density of $n \sim 10^3$\,cm$^{-3}$. The ionization structure and abundances were the subject of a recent study by \cite{MonrealIbero2020}. The nebula is also known to be molecule-rich \cite{Sahai1990}; it is among the brightest PNe in near-IR H$_2$ emission \cite{Storey1984,Kastner1996}. 

A bright A2~V star is found near the centre of the PN, but is too cool to be the ionizing star; the actual PN progenitor is much fainter and is located $\sim$1.7 arcsec to the South-West of the A star \cite{Kohoutek1977,Ciardullo1989}.  
\review{The A2~V star has the same radial velocity and extinction as the PN, and its proper motion ($\mu_\alpha=-7.747$ mas/yr $\sigma_{\mu_\alpha}=0.026$; and
$\mu_\delta=-0.125$ mas/yr  $\sigma_{\mu_\delta}=0.031$) agrees with that of the central star ($\mu_\alpha=-7.677$~mas/yr $\sigma_{\mu_\alpha}=0.235$; and
$\mu_\delta=0.197$ mas/yr $\sigma_{\mu_\delta}=0.275$), demonstrating that the PN progenitor and A-type companion constitute a comoving visual binary. }
The distance to NGC~3132 is obtained from Gaia DR3 measurements of this visual binary. No Gaia DR3 radial velocity is available for the optically faint central star (the PN progenitor). However, the brighter (A-type) visual companion and the PN have the same radial velocity: $(-11.4\pm1.6)$~\kms\ for the A star from Gaia, and $(-10\pm3)$~\kms\ for the PN from \cite{Meatheringham1988}. The A star and PN central star also have compatible Gaia DR3 proper motions (within 1.5$\sigma$).

The brighter, A-type star has a Gaia DR3 geometric distance (median of geometric distance posterior) of 754~pc, with lower and upper 1$\sigma$-like confidence intervals (16th and 87th percentiles of the posterior) of 18~pc and 15~pc respectively \citep{Bailer-Jones2021}.
The fainter central star  has a Gaia DR3 geometric distance  of 2124.7~pc, with lower and upper 1$\sigma$-like bars of 559.1~pc, and 1464.5~pc. The quality flags of the astrometric solution for this star are not optimal, most likely due to the vicinity of the much brighter A-star; in particular, the goodness-of-fit along the scan is 16.9, while it should be close to unity. 
We therefore adopt the Gaia DR3 distance to the central star's visual A-type companion, $754^{+15}_{-18}$\,pc, as the distance to the PN. 

\subsection*{Densities, masses and excitation of the \hh\ knots}
\label{sec:methods-knots}

The clumpiness of NGC~3132 in H$_2$ emission links this nebula to other molecule-rich PNe, such as the Helix Nebula (NGC~7293, \cite{Odell_etal_2004_HelixKnots,Meixner_etal_2005,matsuura_etal_2007,Matsuura_etal_2009}), Ring Nebula (NGC~6720, \cite{Kastner1994}), and the hourglass-shaped (bipolar) nebula NGC~2346 \citep{Manchado2015}, in which the molecular emission seems to be associated with dense knots that are embedded in or surround the ionised gas. The origin of such H$_2$ knots in PNe --- as overdensities in the former AGB wind, vs.\ formation {\it in situ} following recombination of H, as the central star enters the cooling track --- remains an open question \citep{Fang2018}. In contrast to the Helix Nebula, there is little evidence for cometary tails emanating from the knots in the inner regions of NGC~3132. However, NGC~3132's system of approximately radially-directed H$_2$ spikes external to the main H$_2$-bright ring system has close analogues in, e.g., the Ring and Dumbbell Nebulae \cite{Kastner1994,Kastner1996}.

Some H$_2$ knots in NGC~3132 are seen in absorption against the bright background nebular emission.
This extinction is apparent not only in optical (HST) images but also, surprisingly, even in the JWST NIRCam near-infrared images (see Supplementary Figure~4). 
We measured the extinction at 1.87 $\mu$m for two knots seen in absorption against the (Pa$\alpha$) nebula background: the largest knot on the west side (coordinates 10:07:00.4, $-$40:26:08.8), and one of the darkest on the east side (10:07:02.5, $-$40:26:00.3). The diameters of these knots are  $\sim$0.36\,arcsec and $\sim$0.15\,arcsec, while their extinction is $\sim$0.57\,mag and $\sim$0.25\,mag (at 1.87 $\mu$m), respectively; using the dust extinction law $A(\lambda)/A(V)$ from \cite{Cardelli:1989p1977}, the corresponding values of $A(V)$ are 3.9 and 1.7\,mag assuming $R_V=3.1$. We then estimate the hydrogen column densities $N(\rm{H})$ from these extinction measurements, and convert to the hydrogen density $n(\rm{H})$ of the knot by assuming that the knot diameters are roughly equivalent to their depths along the line of sight. Using the conversion between $A(V)$ and $N(\rm{H})$ from \cite{Bohlin.1978}, where H is the combination of H$^0$, H$^+$ and H$_2$, the estimated column densities are $N(\rm{H}) = 7.3\times10^{21}$\,cm$^{-2}$ and $N(\rm{H}) = 3.2\times10^{21}$\,cm$^{-2}$, respectively.
For the adopted distance of 754\,pc, the estimated densities are $n(\rm{H}) \sim 2\times10^6$\,cm$^{-3}$ for both knots. These densities suggest knot masses of $10^{-5}$\,M$_\odot$, similar to the typical knot (``globule'') masses found in the Helix Nebula \citep{Andriantsaralaza2020}. 

The critical density of excitation of the 2.12 \mic\ H$_2$ 1--0 S(1) line at a kinetic temperature of 2000\,K is 9$\times$10$^5$\,cm$^{-3}$ \citep{Bourlot1999}, if the collision partner is H. The critical density is higher for the 1--0 S(1) line than for the 0--0 S(9) 4.69~\mic\ H$_2$ line \citep[6$\times$10$^4$~cm$^{-3}$;][]{Wolniewicz_etal_1998,Bourlot1999}. 
Hence, the excitation of H$_2$ should be nearly thermal if the gas temperature is sufficiently high, with the caveat that both critical densities are higher if the primary collision partner is H$_2$ rather than H.

\subsection*{PSF measurements of the central star}
\label{sec:methods-central-star}

To ascertain whether the mid-IR source associated with the PN central star is extended, we measured the JWST instrumental point spread function (PSF), using Gaussian fitting of field stars. We measured Gaussian FWHMs of 0.29, 0.40,  0.44 and 0.58 arcsec at 7.7, 11.3, 12.8 and 18 \mic, respectively. We also measured two compact, slightly resolved  galaxies in the field. 

We then repeated the procedure for the central star. No fit was possible at 18 \mic, due to saturation (see Supplementary Figure~5). At 7.7 \mic\ the central star is on the edge of the diffraction spike of the A star, and only an upper limit on FWHM could be obtained. However, measurements of the PN central star image in the 11.3 and 12.8 \mic\ filters gave consistent results, with measured FWHMs of 0.55 and 0.60 arcsec, significantly larger than the respective PSFs and comparable to the two field galaxies. Gaussian deconvolution using the PSF yields deconvolved FWHM values for the central star of $\leq 0.3$ ($\leq 230$~AU) at 7~\mic,  and 0.4~arcsec (300~AU) at 11.3 and 12.8~\mic. \review{The extent of the central star at 18~\mic\, is $\gtrsim0.9$~arcsec in diameter (see Supplemenraty Figures~5 and 7).}

\subsection*{Central star system's masses}
\label{sec:methods-central-star-mass}

We determined the mass of the A-star companion using version 1.2 of the PARSEC isochrones \citep{Marigo2017}
for solar metallicity, taken as $Z=0.0152$. We used $M_{\rm bol, 0} = (0.34\pm0.25)$~mag and the GAIA DR3 spectroscopic temperature $T_{\rm eff}=(9200\pm200)$\,K, where the errors are conservative. The star is confirmed to be beginning to turn off the main sequence, in a phase where the luminosity of $(57\pm15)\,\rm L_\odot$ increases by 0.1\%\ per Myr and the temperature decreases by 7\,K per Myr (see Supplementary Figure~6). The isochrones yield an age of $(5.3\pm0.3)\times 10^8\,$yr and a mass of $M_{\rm A2V}=(2.40\pm0.15)\,\rm M_\odot$.  The central star of the PN is evolving on the same isochrone, but from a more massive star as it has evolved further. We use the same isochrones to determine the initial mass of a star on the thermal-pulsing AGB, the phase where the central star ejected the envelope. This gives an initial mass for the central star of $M_{\rm i}= (2.86\pm 0.06)\,\rm M_\odot$.  \review{We have carried out the same isochrone fitting using an alternative stellar evolutionary model (the DARTMOUTH code; \cite{Dotter2008}). Both the A2V star mass and the mass of the progenitor of the central star decrease by 0.15~\Msun.}

\review{The final, CS, mass for such a star is 0.66~\Msun. However, we have shown that such a star would show a high C/O$\sim$2, while the presence of silicate features in the Spitzer spectrum indicate that C/O$\lesssim$1. To reconcile the mass and the abundances we conjecture that the evolution was interrupted by the binary interaction that formed the disk, when the core mass was 0.61~\Msun. With such a mass the evolutionary time to the current position on the HR diagram is in better agreement with the age of the nebula. This mass is also in better agreement with that derived from the photoionisation model (0.58$\pm$0.03)~\Msun.}

\comment{\sout{The elemental abundances of a PN can be used independently to constrain the initial stellar mass by comparing them with the final yields of stellar evolution. Alternatively, if the mass is constrained from an alternative method, as in our case, it can be used to constrain chemical evolution models.
The N/O ratio measured for the gas PN phase (0.42; \cite{MonrealIbero2020}; see Section~\ref{sec:methods-photoionisation-modelling}) indicating that stars of mass 2.86~\Msun\ do not suffer hot bottom burning (HBB). Recent stellar structure calculations  \cite{Ventura2018} show that HBB takes place for (initial) stellar masses  $\gtrsim 3$~\Msun\ (ATON models) or $\gtrsim 4$~\Msun\ (MONASH models); so the first set of models is in reasonable and better agreement with our observational constraint.}}

\comment{\sout{The C/O ratio tends to be lower in stars that suffer HBB, because carbon is converted to nitrogen. The gas phase C/O measurements of the PN are not reliable, but the dust around the central star is likely oxygen, rather than carbon rich \citep{Delgado-Inglada2014}, an indication that the C/O ratio of the gas phase could be below unity. The same models predict that to have a C/O$\lesssim$ 1 the star must have an initial mass of $\gtrsim$3.25~\Msun\ (ATON models) or $\gtrsim$ 4.75~\Msun\ (MONASH models). So, we can preliminarily state that a C/O ratio below unity may be in approximate agreement with the ATON models, although an actual gas phase C/O ratio measurement is needed.}}

\comment{\sout{The N/O abundance ratio from our 3D ionisation model (N/O$\sim$0.4, see Section~\ref{sec:methods-photoionisation-modelling}) compared to the yields of the final thermal pulses at solar metallicity \citep{Ventura2018}}}
\comment{\sout{sets the mass of the central star's progenitor at $\sim$3 M$_{\odot}$, which is consistent with the progenitor mass inferred from the presence of its visual binary companion. Adopting the initial-final mass relations in \cite{Marigo2020}, we predict the central star presently has a mass of $M_f = 0.71\pm0.05\,\rm M_\odot$. This estimate is however dependent on mass-loss models and the impact of their uncertainty on the result is not well quantified.}}

\subsection*{Photoionisation modelling}
\label{sec:methods-photoionisation-modelling}

\review{The stratified ionisation and excitation structure of NGC~3132 is evident in Fig.~\ref{fig:HIIvsH2}, wherein the bright rim of ionized gas, as traced by [\siii] and Br$\alpha$ emission, lies nestled inside the peak \hh\  emission.
However, significant ionised hydrogen and high-excitation plasma — traced by [\neii] and [\siii] emission in the MIRI F1280W and F1800W filter images, respectively — is observed beyond the bright inner, elliptical ring. }

We constructed a three-dimensional photoionisation model using the code Mocassin \cite{Ercolano2003}. To constrain the model we used the Multi Unit Spectroscopic Explorer (MUSE) emission line maps and absolute H$\beta$ flux of \cite{MonrealIbero20}, the optical integrated line fluxes from \cite{Tsamis2004}, the IR line fluxes from \cite{Mata_etal_2016}, as well as the velocity-position data obtained from the high-resolution scanning Fabry-Perot interferometer, SAM-FP,  mounted on the SOAR telescope adaptive module. The observations were taken under photometric conditions. The seeing during the observations was 0.7 arcsec for the [\nii] observations to 0.9~arcsec for the H$\alpha$ one. The FWHM of a Ne calibration lamp lines was 0.586~\AA\ or 26.8~km~s$^{-1}$, which corresponds to a spectral resolution of about 11\,200 at H$\alpha$.

We determined the density structure by fitting the emission line maps to the SAM-FP images of [\nii] $\lambda$6584 and H$\alpha$, using a distance of 754~kpc.
The model adopts as free parameters the temperature and luminosity of the ionising source, and the elemental abundance of the gas component (assumed constant throughout the nebula); we assumed that no dust is mixed in the gas. For the ionising source we use the NLTE model atmospheres of central stars of planetary nebulae from \cite{Rauch2000}. 

We find that a model invoking an unobscured central star with effective temperature $T_\mathrm{eff}=110$~kK and luminosity $L=200$~L$_{\odot}$ well matches the observational data. However, we find that the present-day central star mass implied by the comparison, between these stellar parameters and the evolutionary tracks of 
\review{\cite{Ventura2018}}
\comment{\sout{\citep{Bertolami2016}}} \review{($0.58\pm0.03$~\Msun)}\comment{\sout{(see Figure~\ref{fig:HRD-Z0p02})}} is inconsistent with the (large) initial mass inferred from consideration of the presence of the comoving, wide-separation A-type companion (\review{0.66~\Msun}; see Central star system's masses).
Furthermore, the tracks of\comment{\sout{\cite{Bertolami2016}}} \review{\cite{Ventura2018}} indicate that, for this mass, we would have a post-AGB age of\comment{\sout{48\,000~yrs}} \review{20\,000~yrs}, whereas the position-velocity data from the SAM-FP instrument yield an expansion velocity of 25-35~\kms\, implying a much shorter and inconsistent nebular dynamical age in the range 2200--5700~yrs.

\review{The C/O and N/O abundances of the nebula, as well as the crystalline silicate nature of the dust in the PN,  indicate that this object has not undergone hot bottom burning {\it and} that it has not undergone sufficient dredge up to have increased the C/O ratio above unity. By the time the 2.86-\Msun\ star reaches the tip of the AGB its C/O ratio is approximately 2. It therefore seems that the mass implied by the initial-to-final mass relation using a main sequence mass of 2.86~\Msun, is too high. We have two ways to resolve this inconsistency (which may both be operating). The central star is shielded by dust in the circumstellar disk making it appear, to the PN, as a cooler star, and/or the central star mass is actually smaller than  0.66~\Msun, because the AGB evolution was interrupted by a binary interaction.}

\comment{\sout{Given JWST's detection of thermal IR emission from dust at the central star, we produced a second model that assumes a hotter central star that is partially obscured by circumstellar dust. The estimated dust mass indicates an extinction of $A_V= 0.0002$ mag toward the star.} \review{A simple model where we assume a certain amount of extinction around the central star does not allow us to reproduce the nebular emission line fluxes. This could be due to the simple nature of the model we are using.}}

\review{If the stellar ascent of the AGB was interrupted, we can determine the upper limit for a mass that would produce a nebula with $C/O\lesssim 1$ and N/O$\sim$0.4. This is 0.61~\Msun. The time for a star of this mass to move from the AGB to the location on the HR diagram with an approximate temperature and luminosity (110kK, 200~\Lsun) as  measured above is $\sim$10\,000~yrs.}
\review{The time-scale of the transition from AGB to post-AGB and PN is tightly connected with the rate at which the envelope is consumed: the results obtained are therefore sensitive to the mass-loss description. The time of 10\,000~yr, is based on the classic mass loss rates dictated by Reimers or Blocker \cite{Bloecker1995}. This estimate must be considered as an upper limit of the duration of this phase; indeed the recent works on the AGB to post-AGB transition by \cite{Kamath2021} and \cite{Tosi2022} showed that to reproduce the infrared excess of post-AGB stars in the Galaxy and in the Magellanic Clouds one has to invoke significantly higher mass-loss rates than those based on the aforementioned formulations, something that would reduce the time-scales by a factor of $\sim$5. The timescale of 10\,000~yrs is therefore easily reconciled with the observed timescale of 2200-5700~yrs implied by the nebula.}

\comment{\sout{Adopting this value and $R_V=3.1$, we found that a central source spectrum with $T_\mathrm{eff}=130$~kK and $L=250$~L$_{\odot}$ is able to reproduce reasonably well the observations. As shown in Figure~\ref{fig:HRD-Z0p02}, this dust-obscured central star model has a larger present-day  mass ($\sim$0.6 \Msun), which is in better agreement with the final central star mass (0.71~\Msun) implied by the determined initial mass of $\sim$2.86~\Msun\ using  the A-type companion (Section~\ref{sec:methods-central-star-mass}), while the evolutionary age is more compatible with the nebular dynamical age. }}

\subsection*{Hydrodynamic modelling}
\label{sec:methods-hydro-modelling}

The hydrodynamic simulation used to interpret the  fragmentation and radial spikes is a 2-dimensional hydrodynamic simulation  using the magneto-hydrodynamic code ZEUS-3D. The computational grid is in spherical coordinates and consists of 800 $\times$ 800 equidistant zones in $r$ and 
$\theta$ respectively, with an angular extent of $90^{\circ}$. The wind and UV luminosity inputs correspond 
to a stellar post-AGB model with 0.677 \Msun \,  which evolves from an initial 2.5 \Msun \, main sequence star \cite{Villaver2002}. 

At simulation time 0~yr the star has $T_{\rm eff}=10\,000$~K and the AGB wind ($v=10$~\kms, $\dot{M}=10^{-6}$~\Msun~yr$^{-1}$) has a homogeneous distribution outside of the pre-PN. The pre-PN has had 1000~yr of evolution prior to this moment, during which time a wide magnetic jet operated with a velocity $v = 230$~\kms, and a mass-loss rate $\dot{M} = 1.3\times 10^{-7}$~\Msun~yr$^{-1}$; this simulation is taken  from Model C6 in \cite{GarciaSegura2021}. At this time the star starts emitting a fast tenuous wind with a velocity $v$ from 240 to 14\,000~\kms\ and a mass-loss rate, $\dot{M}$ ranging from 
$1.06\times 10^{-7}$ to $1.13\times 10^{-10}$~\Msun~yr$^{-1}$ over 4000~yrs that sweeps up the AGB wind material. At the same time (0 yr) the ionisation front propagates into the medium. 

\backmatter


\section*{Data availability}
HST data are available at {\it HST} Legacy Archive
(https://hla.stsci.edu). JWST data were obtained from the Mikulski Archive for Space Telescopes at the Space Telescope Science Institute (https://archive.stsci.edu/).
MUSE data were collected at the European Organisation for Astronomical 
Research in the Southern Hemisphere, Chile (ESO Programme 60.A-9100), presented by Monreal-Ibero et al. (2020) are available at the ESO Archive (http://archive.eso.org). San Pedro de Martir data is available at http://kincatpn.astrosen.unam.mx.

\section*{Code availability}
The code MOCASSIN is available at the following URL: https://mocassin.nebulousresearch.org/. ZEUS3-D is available at the Laboratory for Computational Astrophysics  \cite{Clarke1996}).  The compiled version of {\it Shape} is available at http://www.astrosen.unam.mx/shape.

\bmhead{Acknowledgements}

We would like to start by acknowledging the International Astronomical Union that oversees the work of Commission H3 on Planetary Nebulae. It is through the coordinating activity of this committee that this paper came together. 
SA acknowledges support under the grant 5077 financed by IAASARS/NOA. 
JA and VB acknowledge support from EVENTs/NEBULAE WEB research program, Spanish AEI grant PID2019-105203GB-C21. 
IA acknowledges the support of CAPES, Brazil (Finance Code 001). 
EDB acknowledges financial support from the Swedish National Space Agency.
EB acknowledges NSF grants AST-1813298 and PHY-2020249.
JC and EP acknowledge support from an NSERC Discovery Grant. 
GG-S thanks Michael L.\ Norman and the Laboratory for Computational Astrophysics for the use of ZEUS-3D. 
DAGH and AM acknowledge support from the ACIISI, Gobierno de Canarias and the European Regional Development Fund (ERDF) under grant with reference PROID2020010051 as well as from the State Research Agency (AEI) of the Spanish Ministry of Science and Innovation (MICINN) under grant PID2020-115758GB-I00.
JGR acknowledges support from Spanish AEI under Severo Ochoa Centres of Excellence Programme 2020-2023 (CEX2019-000920-S). 
JGR and VGLL acknowledge support from ACIISI and ERDF under grant ProID2021010074. 
DGR acknowledges the CNPq grant 313016/2020-8. 
MAG acknowledges support of grant PGC 2018-102184-B-I00 of the Ministerio de Educaci\'on, Innovaci\'on y Universidades cofunded with FEDER funds and from the State Agency for Research of the Spanish MCIU through the ``Center of Excellence Severo Ochoa" award to the Instituto de Astrof\'\i sica de Andaluc\'\i a (SEV-2017-0709). 
DJ acknowledges support from the Erasmus+ programme of the European Union under grant number 2020-1-CZ01-KA203-078200. 
AK and ZO were supported by the Australian Research Council Centre of Excellence for All Sky Astrophysics in 3 Dimensions (ASTRO 3D), through project number CE170100013. This research is/was supported by an Australian Government Research Training Program (RTP) Scholarship. 
MM and RW acknowledge support from STFC Consolidated grant (2422911). 
CM acknowledges support from UNAM/DGAPA/PAPIIT under grant IN101220. 
SM acknowledges funding from UMiami, the South African National Research Foundation and the University of Cape Town VC2030 Future Leaders Award.
JN acknowledges support from NSF grant AST-2009713.
CMdO acknowledges funding from FAPESP through projects 2017/50277-0, 2019/11910-4 e 2019/26492-3 and CNPq, process number 309209/2019-6. JHK and PMB acknowledge support from NSF grant AST-2206033 and a NRAO Student Observing Support grant to Rochester Institute of Technology. MO was supported by JSPS Grants-in-Aid for Scientific Research(C) (JP19K03914 and 22K03675).

QAP acknowledges support from the HKSAR Research grants council.
Vera C. Rubin Observatory is a Federal project jointly funded by the National Science Foundation (NSF) and the Department of Energy (DOE) Office of Science, with early construction funding received from private donations through the LSST Corporation. The NSF-funded LSST (now Rubin Observatory) Project Office for construction was established as an operating center under the management of the Association of Universities for Research in Astronomy (AURA). The DOE-funded effort to build the Rubin Observatory LSST Camera (LSSTCam) is managed by SLAC National Accelerator Laboratory (SLAC).

AJR was supported by the Australian Research Council through award number FT170100243.
LS acknowledges support from PAPIIT UNAM grant IN110122.
CSC's work is part of I+D+i project PID2019-105203GB-C22 funded by the Spanish MCIN/AEI/10.13039/501100011033.
MSG acknowledges support by the Spanish Ministry of Science and Innovation (MICINN) through projects AxIN (grant AYA2016-78994-P) and EVENTs/Nebulae-Web (grant PID2019-105203GB-C21). RS’s contribution to the research described here was carried out at the Jet Propulsion Laboratory, California Institute of Technology, under a contract with NASA. J.A.T. would like to thank Marcos Moshisnky Fundation (Mexico) and UNAM PAPIIT project IA101622
EV acknowledges support from the "On the rocks II project" funded by the Spanish Ministerio de Ciencia, Innovaci\'on y Universidades under grant PGC2018-101950-B-I00. 
AZ acknowledges support from STFC under grant ST/T000414/1.  

This research made use of Photutils, an Astropy package for detection and photometry of astronomical sources \citep{photutils}, of the Spanish Virtual Observatory (https://svo.cab.inta-csic.es) project funded by MCIN/AEI/10.13039/501100011033/ through grant PID2020-112949GB-I00 and of the computing facilities available at the Laboratory of Computational Astrophysics of the Universidade Federal de Itajub\'{a} (LAC-UNIFEI, which is maintained with grants from CAPES, CNPq and FAPEMIG).

Based on observations made with the NASA/ESA Hubble Space Telescope, and obtained from the Hubble Legacy Archive, which is a collaboration between the Space Telescope Science Institute (STScI/NASA), the Space Telescope European Coordinating Facility (ST-ECF/ESAC/ESA) and the Canadian Astronomy Data Centre (CADC/NRC/CSA). 

The JWST Early Release Observations and associated materials were developed, executed, and compiled by the ERO production team: Hannah Braun, Claire Blome, Matthew Brown, Margaret Carruthers, Dan Coe, Joseph DePasquale, Nestor Espinoza, Macarena Garcia Marin, Karl Gordon, Alaina Henry, Leah Hustak, Andi James, Ann Jenkins, Anton Koekemoer, Stephanie LaMassa, David Law, Alexandra Lockwood, Amaya Moro-Martin, Susan Mullally, Alyssa Pagan, Dani Player, Klaus Pontoppidan, Charles Proffitt, Christine Pulliam, Leah Ramsay, Swara Ravindranath, Neill Reid, Massimo Robberto, Elena Sabbi, Leonardo Ubeda. The EROs were also made possible by the foundational efforts and support from the JWST instruments, STScI planning and scheduling, and Data Management teams.
 
Finally, this work would not have been possible without the collaborative platforms {\it Slack} (slack.com) and {\it Overleaf} (overleaf.com).


\section*{Author contribution}
The following authors have contributed majorly to multiple aspects of the work that lead to this paper, the writing and the formatting of figures: De~Marco (writing, structure, interpretation, synthesis), Aleman (\hh\ interpretation), Balick (processing and interpreting images), García-Segura (2D hydro modelling), Kastner (writing, \hh\ measurements and interpretation), Matsuura (imaging, photometry, \hh\ interpretation), Miszalski (stellar photometry), Mohamed (hydrodynamics of binaries), Monreal-Ibero (MUSE data analysis), Monteiro (photoionisation and morpho-kinematic models), Moraga Baez (JWST image production), Morisset (photoionisation modelling), Sahai (disk model, comparative interpretation), Soker (hydro modelling, interpretation), Stanghellini (distances, abundance interpretation), Steffen (morpho-kinematic models), Walsh (spatially resolved spectroscopy), Zijlstra (disk model, \hh\ measurements, writing, interpretation).

The following authors have contributed key expertise to aspects of this paper: Akashi (hydrodynamic modelling and jet interpretation), Alcolea (CO observations), Akras (\hh\ interpretation), Amram (space-resolved spectroscopy), Blackman (hydrodynamics), Bublitz (HST and radio images of fast evolving PN), Bucciarelli ({\it Gaia} data), Bujarrabal (radio observations, disk observation and interpretation, comparative studies), Chu (disk interpretation), Cami (molecular formation), Corradi (final review, interpretation), García-Hernandez (IR dust/PAH features and abundances), García-Rojas (photoionisation modelling), Gómez-Llanos (photoionisation modelling), Gonçalves (comparative analysis), Guerrero (Xray imaging), Jones (close binaries), Karakas (final review, stellar nucleosynthesis), Manchado (nebular morphology, \hh\ interpretation),  McDonald (photometry modelling), Montez (X-ray and UV imaging), Osborn (binary nucleosynthesis), Otsuka (IR imaging), Parker (morphology), Peeters (nebular spectroscopy, PAHs), Ruiter (binary populations), Sabin (abundances), Sánchez Contreras (radio), Santander-García (nebular evolution), Seitenzahl (star and star nebula association), Speck (dust), Toalá (morphology), Ueta (nebular imaging), Van de Steene (IR observations), Ventura (AGB evolution model).

The following authors contributed by commenting on some aspects of the analysis and manuscript: De Beck, Boffin, Boumis, Chornay, Frank, Kwok, Lykou, Nordhaus, Oliveira, Quint, Quintana-Lacaci, Redman, Villaver, Vlemmings, Wesson, and Van Winckel.

\section*{Competing interest statement}
We declare that no conflict of interest exists between any of the authors and the content and production of this paper.

\newpage

\section*{Figures}

\begin{figure*}[htbp]
	\includegraphics[width=10cm,angle=90]{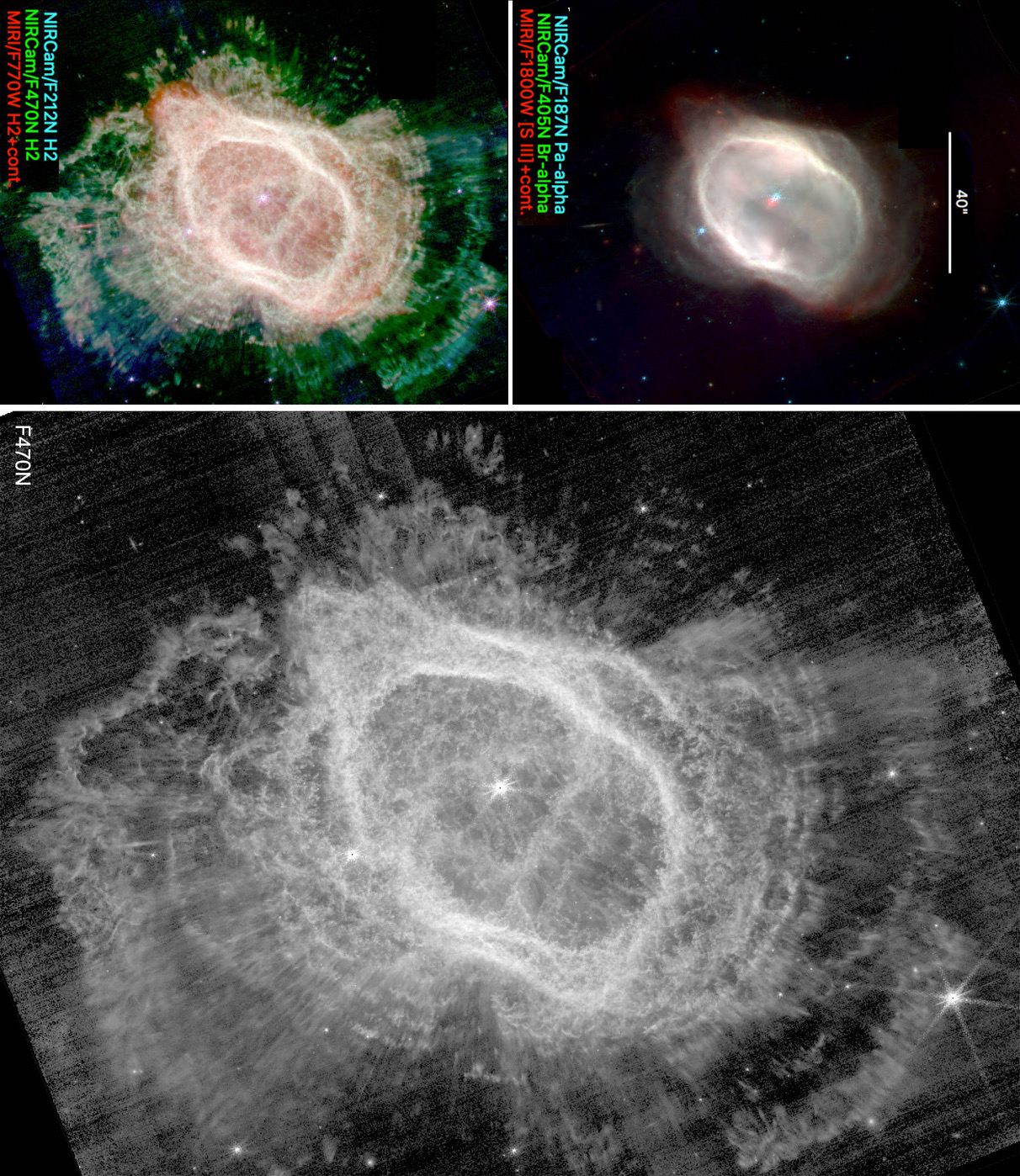}
    \caption{JWST images of the PN NGC~3132. Left column, top and bottom: color overlays of JWST NIRCam and MIRI images that cleanly distinguish between the PN's ionized gas (i.e., H~{\sc ii} region; top panel) and molecular gas (as seen in H$_2$; bottom panel). Note the sharp contrast between the relatively smooth appearance of the H~{\sc ii} region and the flocculent structure of the H$_2$ ring system and extended H$_2$ halo. These images are presented with square-root and log intensity stretches, respectively, from the background sky to peak intensity levels in each image. Right image: a grey-scale, single filter (F470N), zoomed-in NIRCam image that more readily displays details of the flocculent \hh\ halo. North is towards the top, East is towards the left.
    }
    \label{fig:HIIvsH2}
\end{figure*}

\begin{figure*}[htbp]
    \centering
	\includegraphics[width=10cm]{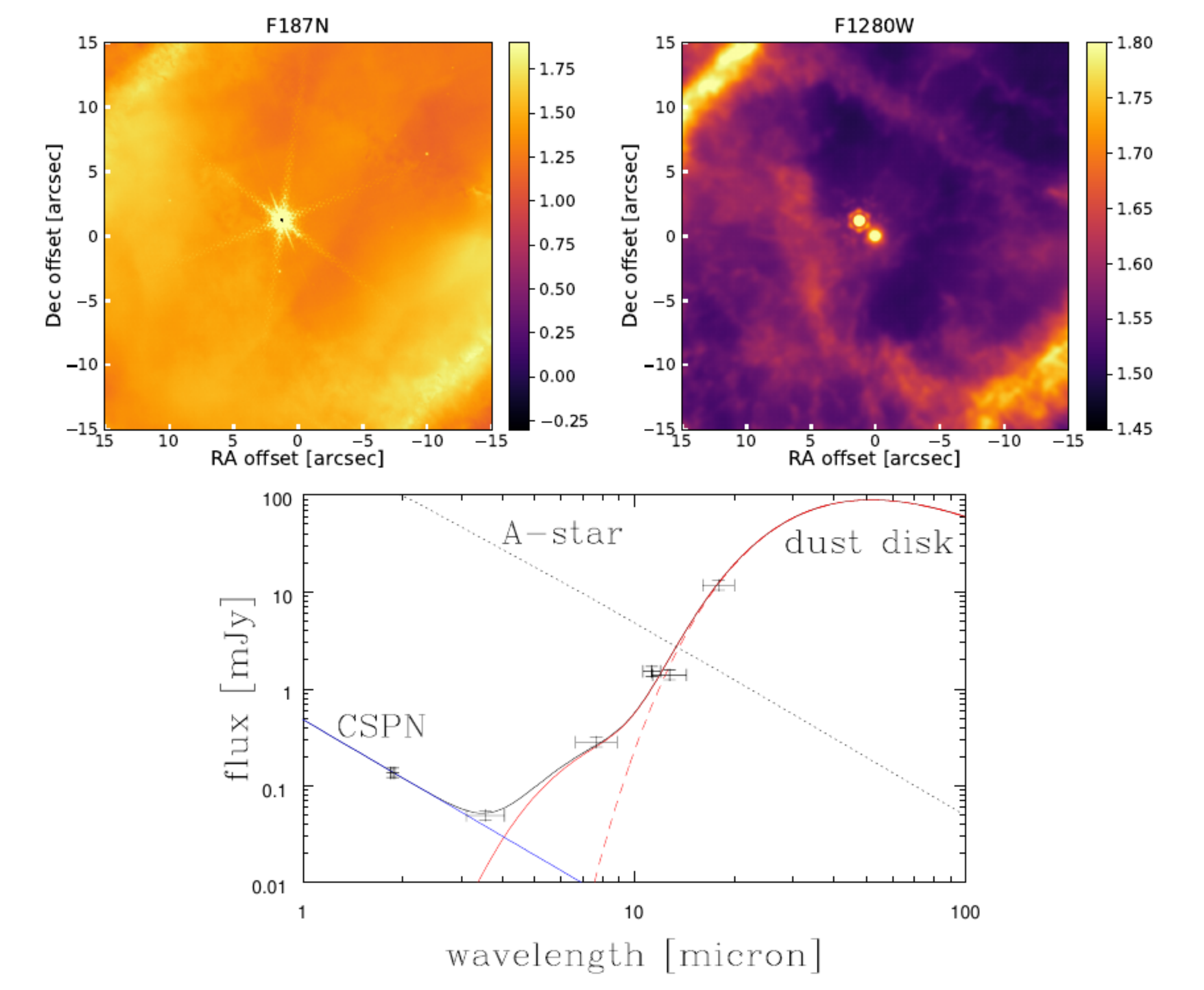}
    \caption{The dusty central star of the PN NGC~3132. JWST NIRCam F187N (top left) and MIRI F1280W (top right) images of the central region of NGC~3132. The JWST MIRI images reveal the detection of a mid-IR excess at the nebula's true (hot, compact) central star, which is seen projected $\sim$1.7$''$ ($\sim$1300AU) SW of the main-sequence A-type companion (which is far brighter shortward of $\sim$10 \mic). North up and East is to the left. Colour bars indicate surface brightness in log (MJy~ster$^{-1}$).  The bottom panel shows the near-IR to mid-IR spectral energy distribution of the central star of NGC~3132  overlaid with a model consisting of a combination of a hot blackbody spectrum representing the central star's photosphere (blue line) and a dusty circumstellar \review{double} disk model to fit the NIR and MIR data points (red line, with the cooler disk as a dashed line). The wide companion, A star's spectral energy distribution is shown as a dotted line. Vertical error bars are set at 10\% of the flux values, while horizzontal bars show the width of the bandpass.}
    \label{fig:CSPNimages_IRexcess}
\end{figure*}

\begin{figure*}[htbp]
    \centering
	\includegraphics[width=12cm]{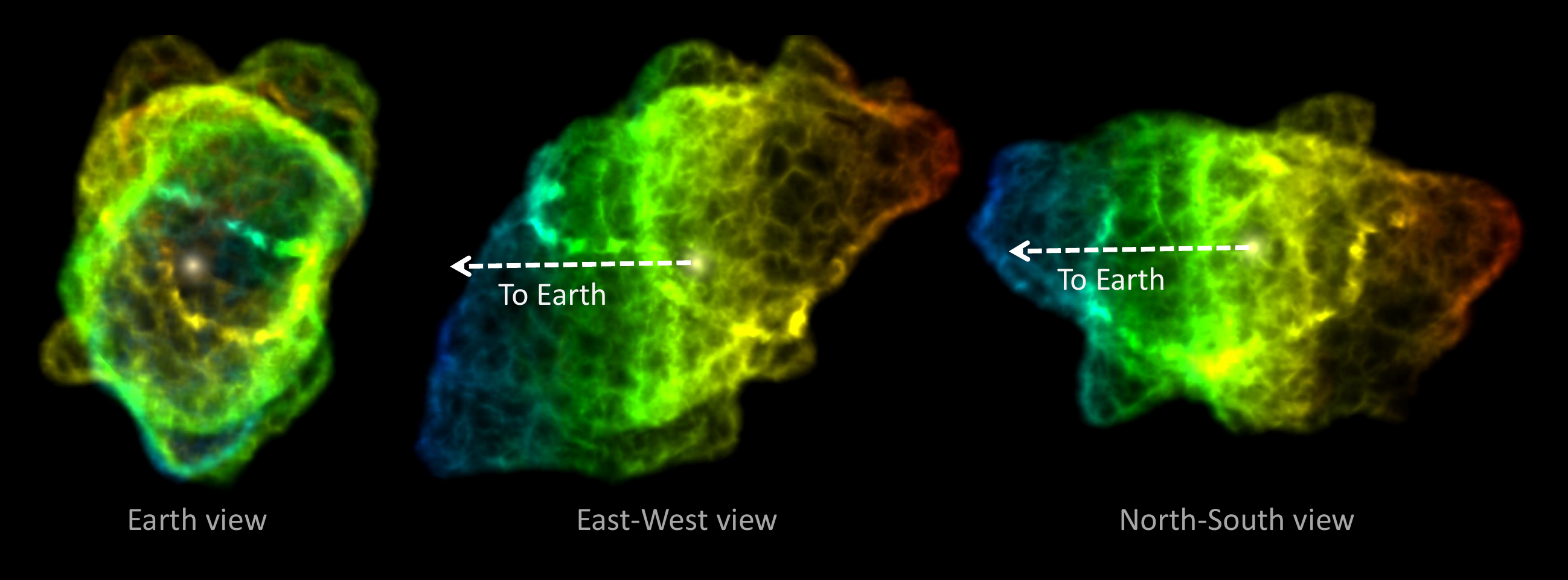}
  \caption{Morpho-kinematic reconstruction of the ionised cavity of PN NGC~3132. Emission in the [\nii] line as seen from Earth (left image; North is towards the top and East is towards the left), a view from the East, which we call East-West view (middle image), and a view from the North which we call North-South view (right image). The colour-coding is for Doppler-shift as seen from Earth, with blue for material approaching the observer, red for receding gas and green for no velocity along the observer's line of sight. We note the prominent green (zero Doppler shift) belt in the middle image, and the filament that wraps around the waist of the ellipsoid and which is red-shifted on one side and blue-shifted on the other. A fly-through movie of this model can be found at this \href{www.ilumbra.com/public/science/ilumbra_NGC3132_NII_reconstruction.mp4}{link}.}
  \label{fig:de-project_shape_views}
\end{figure*}

\begin{figure}[htbp]
	\includegraphics[width=\columnwidth]{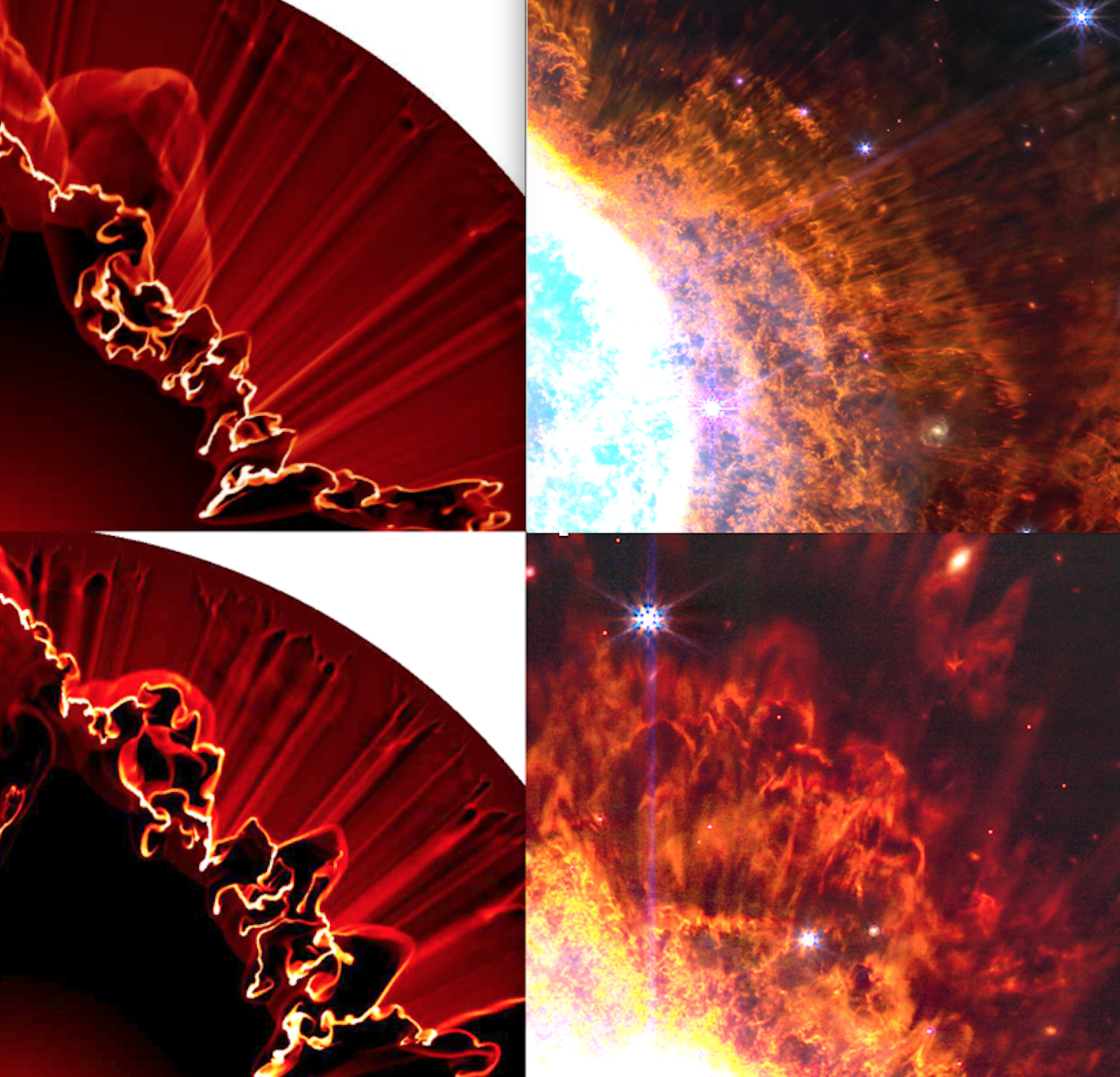}
    \caption{The physical interpretation of the flocculent \hh\ structure. Left column: a hydrodynamic simulation showing the formation of nebular structures external to the main ionised region, compared with (right column) two quadrants of the JWST images of NGC~3132. Top row: the simulation snapshot at 3800~yr from the on-set of ionisation is compared with similar straight spikes in one region of the nebula (top right image North is to the left, East is towards the top) while, bottom row, at 4000~yr the spikes thicken and bend as is also seen in a different part of the nebula (North to the top, and East to the left). This demonstrates temporal evolution in different parts of the nebula.}
    \label{fig:guille-fragmentation2}
\end{figure}

\begin{figure*}[htbp]
    \centering
    \includegraphics[width=12cm]{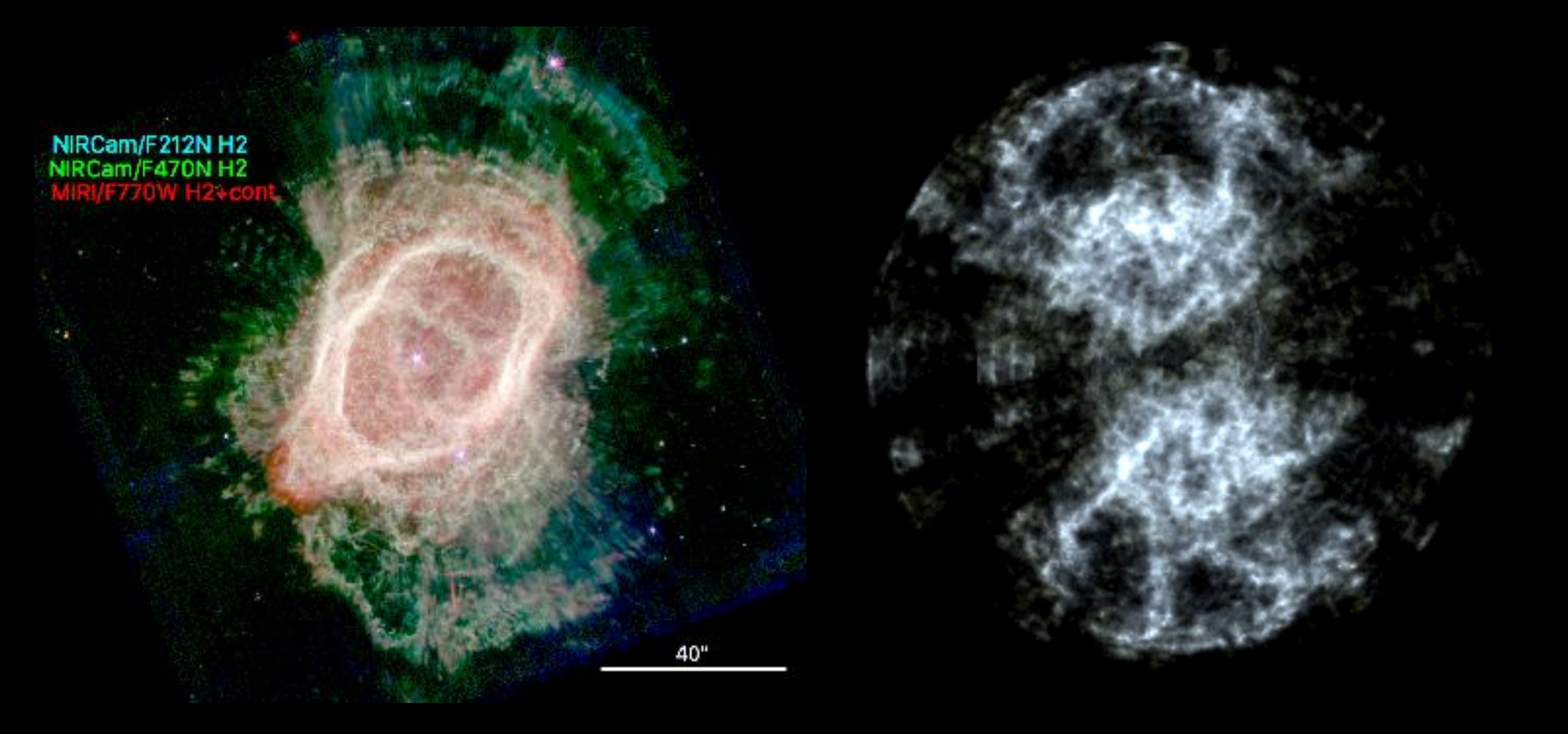}
    \caption{Approximate illumination model of the \hh\ halo of PN NGC~3132. Left panel: the JWST colour composite image showing the \hh\ extended structure. Right panel the projected {\it Shape} image after assuming two concentric, thick uninterrupted shells of material, illuminated by the central star, through a porous ellipsoid with reduced opacity in the polar regions. Fly through movie of this model can be found at this  \href{https://www.ilumbra.com/public/science/H2_Shape_scattering.mp4}{link}. }
    \label{fig:h2-and-comparison}
\end{figure*}





\end{document}


\title[JWST PN]%
{The messy death of a multiple star system and the resulting planetary nebula as observed by JWST}




\author*[1,2]{\fnm{Orsola} \sur{De Marco}}\email{orsola.demarco@mq.edu.au}

\author[3,4]{\fnm{Muhammad} \sur{Akashi}}

\author[5]{Stavros Akras}

\author[6]{Javier Alcolea} 

\author[7]{Isabel Aleman} 

\author[8]{Philippe Amram} 

\author[9]{Bruce Balick} 

\author[10]{Elvire De Beck} 

\author[11,12]{Eric G. Blackman}

\author[13]{Henri M. J. Boffin}

\author[5]{Panos Boumis} 

\author[14]{Jesse Bublitz} 

\author[15]{Beatrice Bucciarelli}

\author[6]{Valentin Bujarrabal} 

\author[16,17,18]{Jan Cami}

\author[19]{Nicholas Chornay}

\author[20]{You-Hua Chu}

\author[21,22]{Romano L.M. Corradi} 

\author[23]{Adam Frank}

\author[24]{Guillermo Garc\'{i}a-Segura}

\author[22,25]{D. A. Garc\'{i}a-Hern\'{a}ndez}

\author[22,25]{Jorge Garc\'{i}a-Rojas}

\author[22,25]{Veronica G\'{o}mez-Llanos}

\author[26]{Denise R. Gon\c{c}alves}

\author[27]{Mart\'{i}n A. Guerrero} 

\author[22,25]{David Jones} 

\author[28,29]{Amanda I. Karakas} 

\author[30,31]{Joel H. Kastner} 

\author[32]{Sun Kwok} 

\author[33,34]{Foteini Lykou} 

\author[22,25,35]{Arturo Manchado}

\author[36]{Mikako Matsuura}

\author[37,38]{Iain McDonald}

\author[39]{Ana Monreal-Ibero}

\author[7]{Hektor Monteiro}

\author[31]{Paula Moraga Baez}  

\author[24]{Christophe Morisset}

\author[40]{Brent Miszalski} 

\author[41,42,43,44]{Shazrene S. Mohamed}

\author[45]{Rodolfo Montez Jr.} 

\author[46,47]{Jason Nordhaus}

\author[48]{Claudia Mendes de Oliveira}

\author[28,29]{Zara Osborn} 

\author[49]{Masaaki Otsuka}

\author[50,51]{Quentin A. Parker}

\author[16,17,18]{Els Peeters}

\author[52]{Bruno C. Quint}

\author[53]{Guillermo Quintana-Lacaci}

\author[54]{Matt Redman}

\author[55]{Ashley J. Ruiter}

\author[24]{Laurence Sabin}

\author[56]{Carmen S\'{a}nchez Contreras}

\author[6]{Miguel Santander-Garc\'{i}a}

\author[55]{Ivo Seitenzahl}

\author[57]{Raghvendra Sahai}

\author[3]{Noam Soker}

\author[58]{Angela K. Speck}

\author[59]{Letizia Stanghellini}

\author[60]{Wolfgang Steffen}

\author[61]{Jes\'{u}s A. Toal\'{a}}

\author[62]{Toshiya Ueta} 

\author[63]{Griet Van de Steene}

\author[56]{Eva Villaver}

\author[64]{Paolo Ventura}

\author[65]{Wouter Vlemmings}

\author[13]{Jeremy R. Walsh}

\author[36]{Roger Wesson}

\author[66]{Hans van Winckel}

\author[38]{Albert A. Zijlstra}


\affil*[1]{School of Mathematical and Physical Sciences, Macquarie University, Sydney, NSW 2109, Australia} 

\affil*[2]{Astronomy, Astrophysics and Astrophotonics Research Centre, Macquarie University, Sydney, NSW 2109, Australia} 

\affil[3]{Department of Physics, Technion, Haifa, 3200003, Israel} 

\affil[4]{Kinneret College on the Sea of Galilee, Samakh 15132, Israel} 

\affil[5]{Institute for Astronomy, Astrophysics, Space Applications and Remote Sensing, National Observatory of Athens, GR 15236 Penteli, Greece} 

\affil[6]{Observatorio Astron\'{o}mico Nacional (OAN/IGN), Alfonso XII, 3, 28014 Madrid, Spain} 

\affil[7]{Instituto de F\'{i}sica e Qu\'{i}mica, Universidade Federal de Itajub\'{a}, Av. BPS 1303, Pinheirinho, Itajub\'{a} 37500-903, Brazil} 

\affil[8]{Aix-Marseille Univ., CNRS, CNES, LAM (Laboratoire d’Astrophysique de Marseille), Marseille, France} 

\affil[9]{Astronomy Department, University of Washington, Seattle, WA 98105-1580, USA} 

\affil[10]{Department of Space, Earth and Environment, Chalmers University of Technology, S-41296 Gothenburg, Sweden} 

\affil[11]{Department of Physics and Astronomy, University of Rochester, Rochester, NY 14627, USA} 

\affil[12]{Laboratory for Laser Energetics, University of Rochester, Rochester NY, 14623, USA} 

\affil[13]{European Southern Observatory, Karl-Schwarzschild Strasse 2, D-85748 Garching, Germany} 

\affil[14]{Green Bank Observatory, 155 Observatory Road, PO Box 2, Green Bank, WV 24944, USA} 

\affil[15]{INAF - Osservatorio Astrofisico di Torino, Via Osservatorio 20, 10023, Pino Torinese, Italy} 

\affil[16]{Department of Physics \& Astronomy, University of Western Ontario, London, ON, N6A 3K7, Canada}

\affil[17]{Institute for Earth and Space Exploration, University of Western Ontario, London, ON, N6A 3K7, Canada}

\affil[18]{SETI Institute, 399 Bernardo Avenue, Suite 200, Mountain View, CA 94043, USA}

\affil[19]{Institute of Astronomy, University of Cambridge, Madingley Road, Cambridge CB3 0HA, UK} 

\affil[20]{Institute of Astronomy and Astrophysics, Academia Sinica (ASIAA), No. 1, Section 4, Roosevelt Road, Taipei 10617, Taiwan}

\affil[21]{GRANTECAN, Cuesta de San Jos\'{e} s/n, E-38712, Bre\~{n}a Baja, La Palma, Spain} 

\affil[22]{Instituto de Astrof\'isica de Canarias, E-38205 La Laguna, Tenerife, Spain} 

\affil[23]{Department of Physics and Astronomy, University of Rochester, Rochester, NY 14627, USA} 

\affil[24]{Instituto de Astronom\'{i}a, Universidad Nacional Aut\'{o}noma de M\'{e}xico, Km. 107 Carr. Tijuana-Ensenada, 22860, Ensenada, B.~C., Mexico} 

\affil[25]{Departamento de Astrof\'{i}sica, Universidad de La Laguna, E-38206 La Laguna, Tenerife, Spain} 

\affil[26]{Observat\'{o}rio do Valongo, Universidade Federal do Rio de Janeiro, Ladeira Pedro Antonio 43, Rio de Janeiro 20080-090, Brazil}

\affil[27]{Instituto de Astrof\'{i}sica de Andaluc\'{i}a, IAA-CSIC, Glorieta de la Astronom\'{i}a, s/n, E-18008, Granada, Spain} 

\affil[28]{School of Physics \& Astronomy, Monash University, Clayton VIC 3800, Australia} 

\affil[29]{ARC Centre of Excellence for All Sky Astrophysics in 3 Dimensions (ASTRO 3D)} 

\affil[30]{Center for Imaging Science, Rochester Institute of Technology, Rochester, NY 14623, USA} 

\affil[31]{School of Physics and Astronomy and Laboratory for Multiwavelength Astrophysics, Rochester Institute of Technology, USA} 

\affil[32]{Department of Earth, Ocean, and Atmospheric Sciences, University of British Columbia, Vancouver, Canada} 

\affil[33]{Konkoly Observatory, Research Centre for Astronomy and Earth Sciences, E\"otv\"os Lor\'and Research Network (ELKH), Konkoly-Thege Mikl\'os \'ut 15-17, 1121 Budapest, Hungary} 

\affil[34]{CSFK, MTA Centre of Excellence, Konkoly-Thege Mikl\'os \'ut 15-17, 1121 Budapest, Hungary} 

\affil[35]{Consejo Superior de Investigaciones Cient\'{\i}ficas, Spain} 

\affil[36]{School of Physics and Astronomy, Cardiff University, The Parade, Cardiff CF24 3AA, UK} 

\affil[37]{Department of Physical Sciences, The Open University, Walton Hall, Milton Keynes, MK7 6AA, UK} 

\affil[38]{Jodrell Bank Centre for Astrophysics, Department of Physics and Astronomy, The University of Manchester, Oxford Road M13 9PL Manchester, UK} 

\affil[39]{Leiden Observatory, Leiden University, Niels Bohrweg 2, NL 2333 CA Leiden, The Netherlands} 

\affil[40]{Australian Astronomical Optics, Faculty of Science and Engineering, Macquarie University, North Ryde, NSW 2113, Australia} 

\affil[41]{Department of Physics, University of Miami, Coral Gables, FL 33124, USA}

\affil[42]{South African Astronomical Observatory, P.O. Box 9, 7935 Observatory, South Africa}

\affil[43]{Astronomy Department, University of Cape Town, 7701 Rondebosch, South Africa}

\affil[44]{NITheCS National Institute for Theoretical and Computational Sciences, South Africa}

\affil[45]{Center for Astrophysics, Harvard \& Smithsonian, 60 Garden Street, Cambridge, MA 02138, USA} 

\affil[46]{Center for Computational Relativity and Gravitation, Rochester Institute of Technology, Rochester, NY 14623, USA} 

\affil[47]{National Technical Institute for the Deaf, Rochester Institute of Technology, Rochester, NY 14623, USA} 

\affil[48]{Departamento de Astronomia, Instituto de Astronomia, Geof\'isica e Ci\^encias Atmosf\'ericas da USP, Cidade Universit\'aria, 05508-900, S\~ao Paulo, SP, Brazil} 

\affil[49]{Okayama Observatory, Kyoto University, Honjo, Kamogata, Asakuchi, Okayama, 719-0232, Japan}

\affil[50]{Department of Physics, CYM Physics Building, The University of Hong Kong, Pokfulam, Hong Kong SAR, PRC} 

\affil[51]{Laboratory for Space Research, Cyberport 4, Cyberport, Hong Kong SAR, PRC} 

\affil[52]{Rubin Observatory Project Office, 950 N. Cherry Ave., Tucson, AZ 85719, USA} 

\affil[53]{Dept. of Molecular Astrophysics. IFF-CSIC. C/ Serrano 123, E-28006, Madrid, Spain} 

\affil[54]{Centre for Astronomy, School of Physics, National University of Ireland Galway, Galway H91 CF50, Ireland}

\affil[55]{University of New South Wales, Australian Defence Force Academy, Canberra, Australian Capital Territory, Australia}

\affil[56]{Centro de Astrobiolog\'{i}a (CAB), CSIC-INTA, Camino Bajo del Castillo s/n, ESAC campus, 28692, Villanueva de la Ca\~{n}ada, Madrid, Spain} 

\affil[57]{Jet Propulsion Laboratory, California Institute of Technology, CA 91109, Pasadena, USA} 

\affil[58]{University of Texas at San Antonio, Department of Physics and Astronomy, Applied Engineering and Technology Building, One UTSA Circle, San Antonio, TX 78249, United States}

\affil[59]{NSF's NOIRLab, 950 N. Cherry Ave., Tucson, AZ 85719, USA} 

\affil[60]{ilumbra, AstroPhysical MediaStudio, Hautzenbergstrasse 1, 67661 Kaiserslautern, Germany} 

\affil[61]{Instituto de Radioastronom\'{i}a y Astrof\'{i}sica, UNAM, Antigua Carretera a P\'{a}tzcuaro 8701, Ex-Hda. San Jos\'{e} de la Huerta, Morelia 58089, Mich., Mexico} 

\affil[62]{Department of Physics and Astronomy, University of Denver, 2112 E Wesley Ave., Denver, CO 80208, USA} 

\affil[63]{Royal Observatory of Belgium, Astronomy and Astrophysics, Ringlaan 3, 1180 Brussels, Belgium} 

\affil[64]{INAF -- Osservatorio Astronomico di Roma, Via Frascati 33, I-00040, Monte Porzio Catone (RM), Italy}

\affil[65]{Onsala Space Observatory, Department of Space, Earth and Environment, Chalmers University of Technology, Onsala, Sweden} 

\affil[66]{Institute of Astronomy, KULeuven, Celestijnenlaan 200D, B-3001 Leuven, Belgium} 

\begin{appendices}

\section*{Supplementary Material}
\label{secA1}

\subsection*{Specifications of JWST NIRCam and MIRI imaging}
\label{sec:supplementary-images}

As part of its ERO program \citep{JWST_ERO}, JWST obtained ten images of NGC~3132: six individual NIRCam images, through filters F090W, F187N, F212N, F356W, F444W, and F470N, and four MIRI images, through filters F770W, F1130W, F1280W, F1800W (Supplementary Figure~\ref{fig:postageStamps}). Basic information about these NIRCam and MIRI filters is presented in Supplementary Table~\ref{tab:filters}. The native NIRCam field of view is $2.2\times 2.2$~arcmin, with a pixel scale of 0.031~arcsec/pixel in the range 0.6–2.3~\mic\ and 0.063~arcsec/pixel in the range 2.4–5.0~\mic; the native MIRI field of view is $1.7 \times 1.3$~arcmin with a pixel scale of 0.11~arcsec/pixel in the range 5-27~\mic. 
The NIRCam instrument provides Nyquist-sampled imaging at 2 (short wavelength channel) and 4 (long wavelength channel) microns with a PSF FWHM of $\sim$2 pixels in both cases.
The MIRI instrument in imaging mode, on the other hand, provides a FWHM of 0.22~arcsec (PSF FWHM of 2 pixels) for wavelengths $\geq$6.25~\mic~ \citep{Bouchet2015}. The NIRCam imaging of NGC~3132 used 8 dither points with an offset of approximately 6 arcsec, while MIRI imaging used a $1\times2$ tile mosaic with 8 dither points.
The final NIRCam and MIRI images cover areas of approximately 150 $\times$ 150 arcsec$^2$ and 150 $\times$130 arcsec$^2$, respectively.

Images \review{were downloaded from the MAST archive (calibration CRDS\_VER11.16.3)} and are neither continuum subtracted (free-free emission is included) nor background subtracted; the background regions of the MIRI images display significant flux that is thermal emission from the (cold) telescope and sunshade. In addition, although specific lines are targeted by specific narrow-band filters, additional (albeit weaker) lines may be present in some bandpasses (e.g., the F187N bandpass, which is dominated by Pa$\alpha$, is contaminated by He\,{\sc I} lines). In order to determine the emission features present in each band, we have generated a simple model to predict the IR spectrum of the nebula, and compared it to previously published {\it Spitzer} observations, see Supplementary Figure~\ref{fig:bandpass-contamination}, alongside the JWST bandpasses. Here we see, for example, that the MIRI F770W image is likely contaminated with \hi\ and [\arii] in certain regions. \review{{\it Spitzer} spectroscopy indicates no sign of PAHs at 6.2, 7.7 and 8.6~\mic\ and only a very weak 11.3~\mic\ feature as well as cristalline silicates \cite{Delgado-Inglada2014,Mata2016}. This is usually associated to the presence of neutral PAHs \cite{Cox2016}. As a result we have not included PAH emission contribution at 7.7~\mic. On the other hand, {\it Spitzer} spectra do not cover the region below 5~\mic\ so we have indicated that there could be a weak 3.3~\mic\ feature. }

In Supplementary Figure~\ref{fig:MikakoH2} we select three regions of the nebula observed through the NIRCam filter F212N, which we present, enlarged, in Supplementary Figure~\ref{fig:MikakoRegions}. The latter Figure presents image sequences consisting of archival HST images and the new JWST (ERO) images. These sequences illustrate the contrast between the smoothness of the emission from ionised gas vs.\ the clumpiness of \hh\ emission, as well as the correspondence between dust extinction (most evident in the HST images) and the \hh\ filaments and knots. 

\begin{table}[htbp]
\begin{center}
\begin{minipage}{250pt}
\caption{List of JWST filters used in this work and expected emission in the corresponding band.} \label{tab:filters}%
\begin{tabular}{lrrl c l}
\toprule
    Filter name & $\lambda_1$\footnotemark[1] & $\lambda_2$\footnotemark[1] & Emission features & Date\footnotemark[2] & $T_\mathrm{exp}$\footnotemark[2]\\
                & ($\mu$m) & ($\mu$m) & within bandpass & & (sec)\\
    \hline
    \multicolumn{6}{c}{NIRCam} \\
    \hline 
    F090W & 0.795 &	1.005 & [\siii] 9069,9562 \AA & 2022-06-03 & 5841\\
    F187N & 1.863 &	1.884 & H {\sc i} Pa$\alpha$ & 2022-06-03 & 9277\\
    F212N & 2.109 &	2.134 & H$_2$ (1,0) $S(1)$ & 2022-06-03& 9277 \\
    F356W & 3.140 &	3.980 & H$_2$; Dust; PAHs? & 2022-06-03 & 1460\\ 
    F405N\footnotemark[3] & 4.028 &	4.074 & H {\sc i}  Br$\alpha$\\
    F470N\footnotemark[3] & 4.683 &	4.733 & H$_2$ (0,0) $S(9)$\\
    F444W & 3.880 & 4.986 & H {\sc i}  Br$\alpha$; H$_2$; Dust? & 2022-06-03 &  2319$\times$2 \\
    \hline
    \multicolumn{6}{c}{MIRI}\\
    \hline
    F770W & 6.6 &	8.8 &  H$_2$ (0,0) $S(5)$ & 2022-06-12 & 2708\\ 
    F1130W & 10.95 & 11.65 & PAHs & 2022-06-12 & 2708\\
    F1280W & 11.6 &	14.0 & [\neii] 12.8$\mu$m;  & 2022-06-12& 2708\\ 
    &&&H$_2$ (0,0) $S(2)$&&\\
    F1800W & 16.5 &	19.5 & Warm dust;  & 2022-06-12 & 2708\\
    &&&[\siii] 18.6$\mu$m&&\\
    \botrule
\end{tabular}
\footnotetext[1]{$\lambda_1$, $\lambda_2$: wavelengths at which the bandpass transmission is 50\% of the peak transmission.} 
\footnotetext[2]{Observing date and exposure time.}
\footnotetext[3]{Pupil wheel filter; used in combination with F444W.}
\end{minipage}
\end{center}
\end{table}

\begin{figure*}[htbp]
	\includegraphics[width=3.9cm]{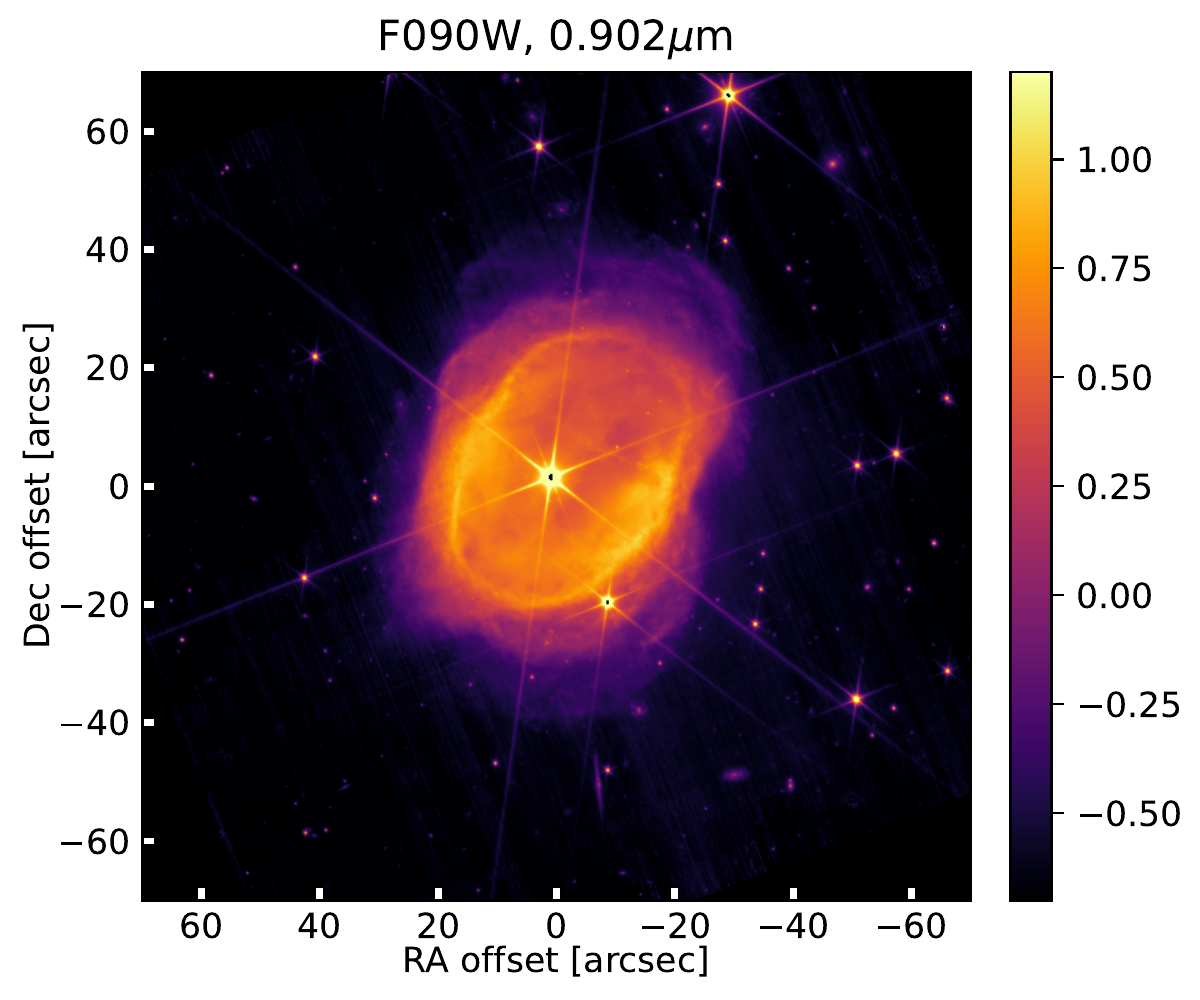}
	\includegraphics[width=3.9cm]{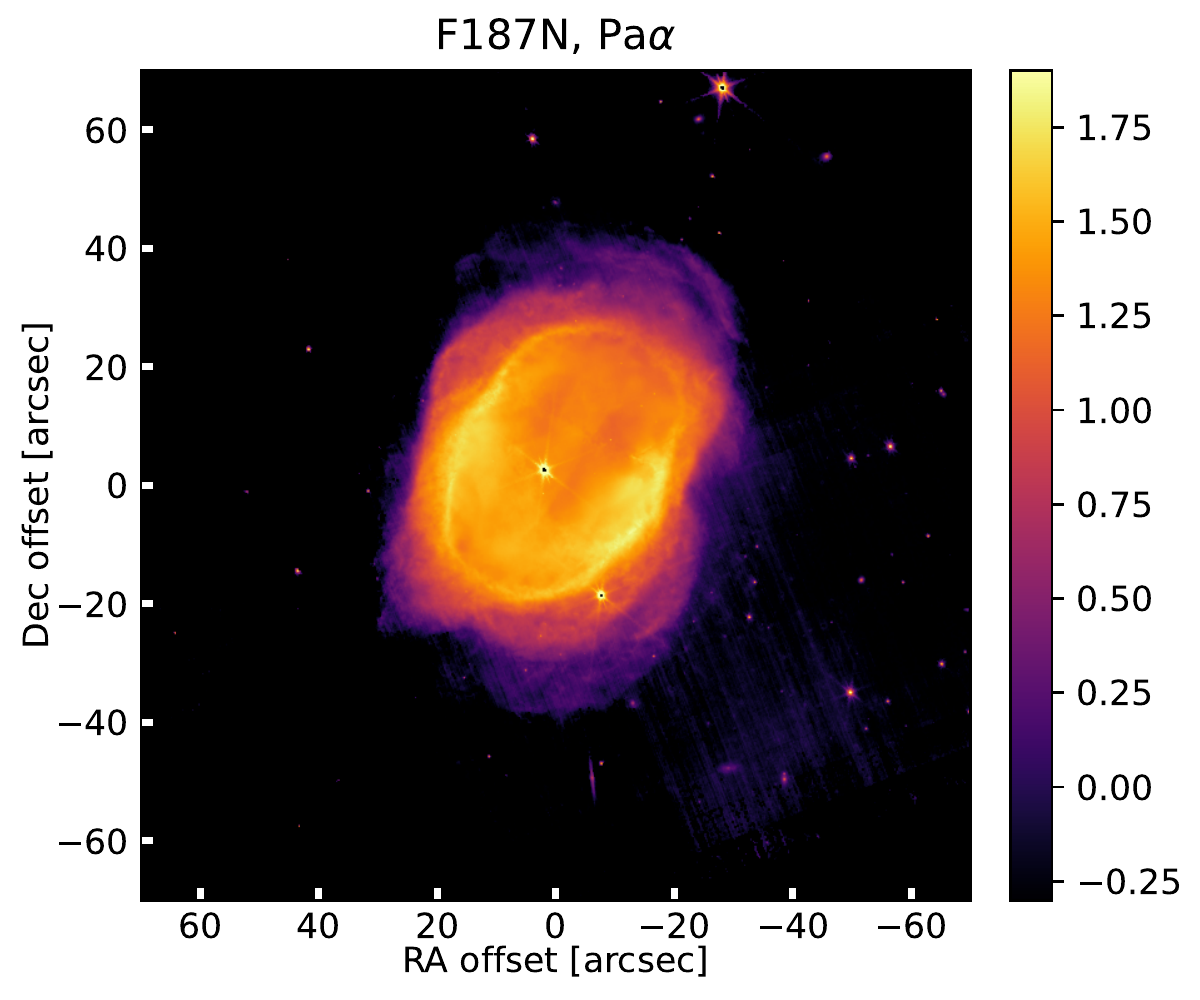}
	\includegraphics[width=3.9cm]{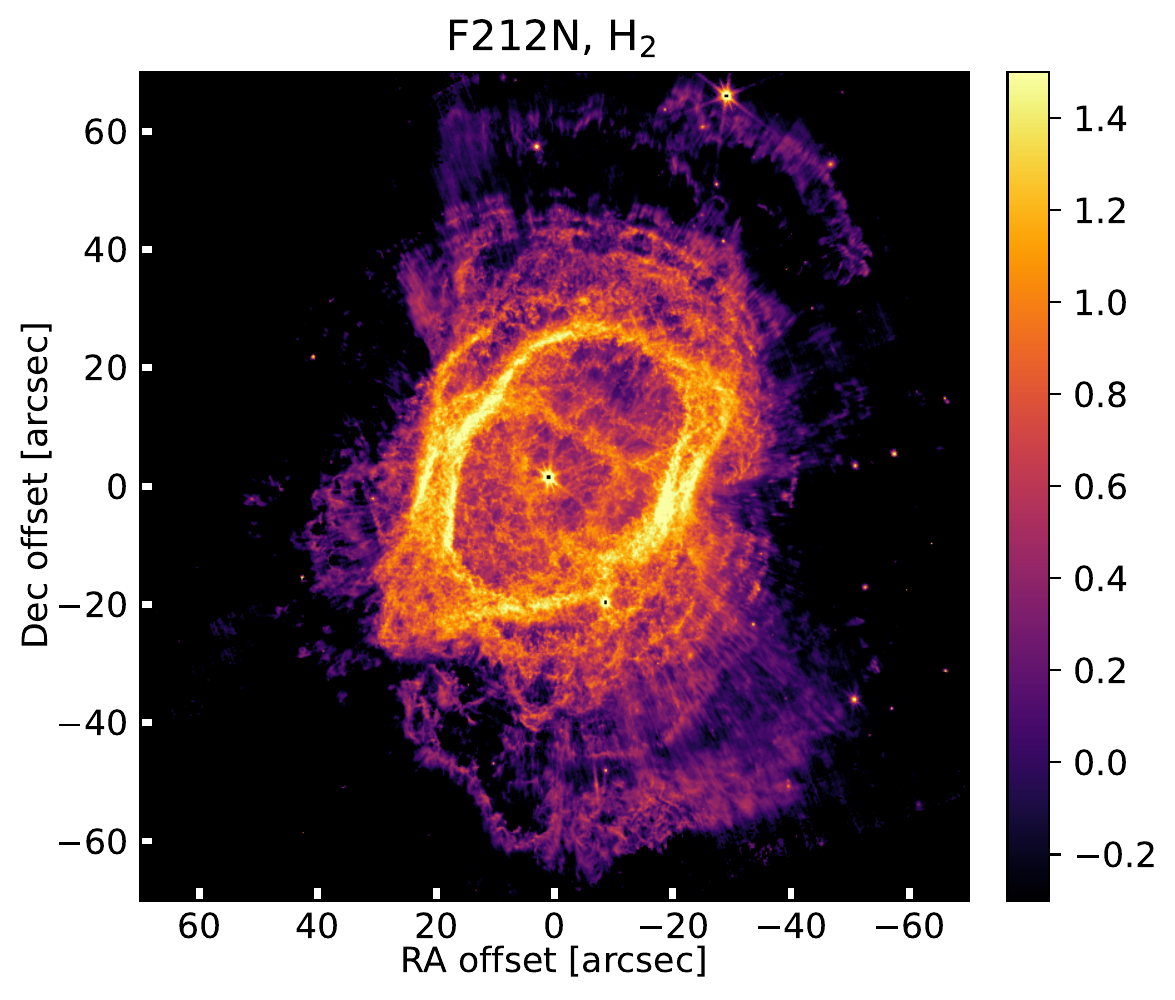}
	\includegraphics[width=3.9cm]{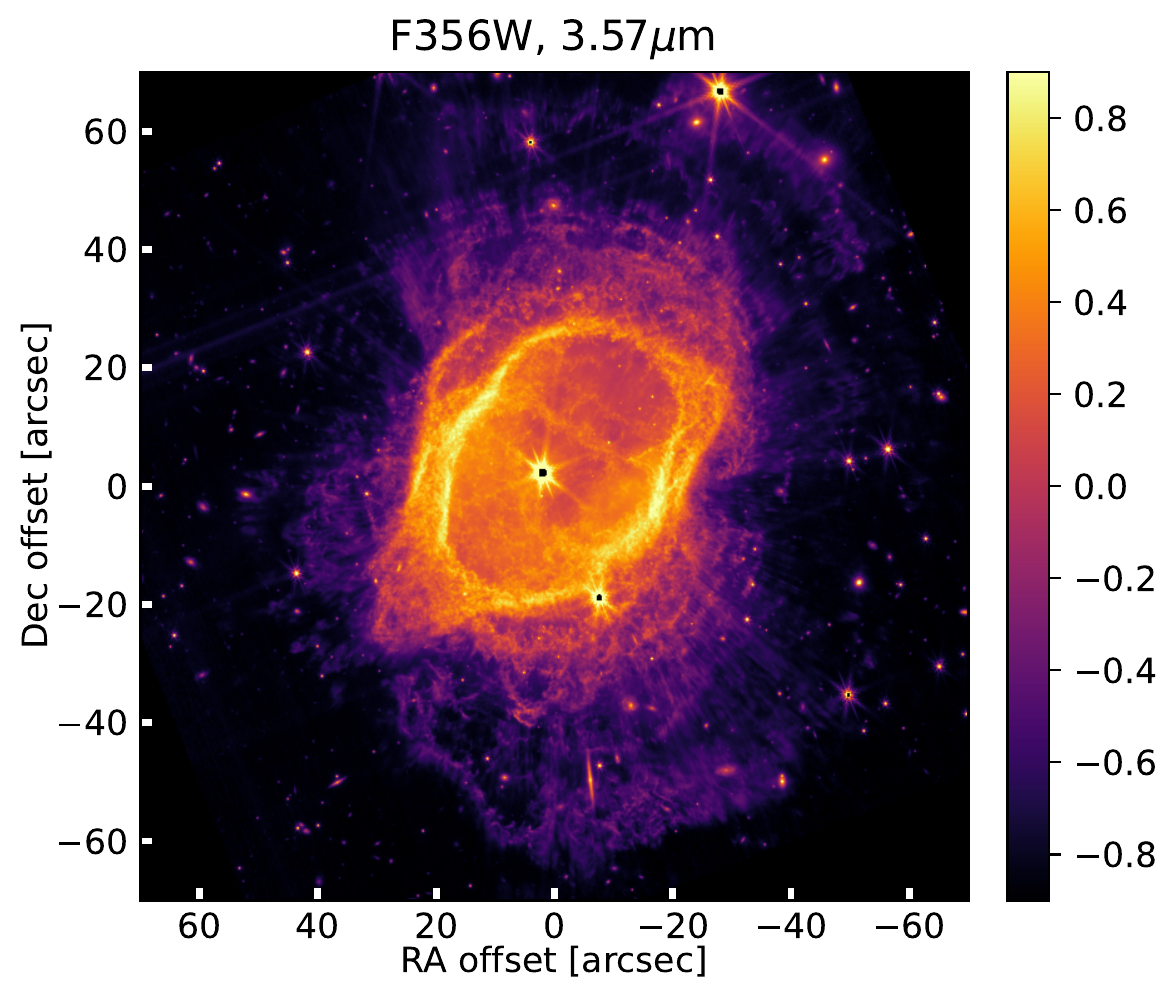}
	\includegraphics[width=3.9cm]{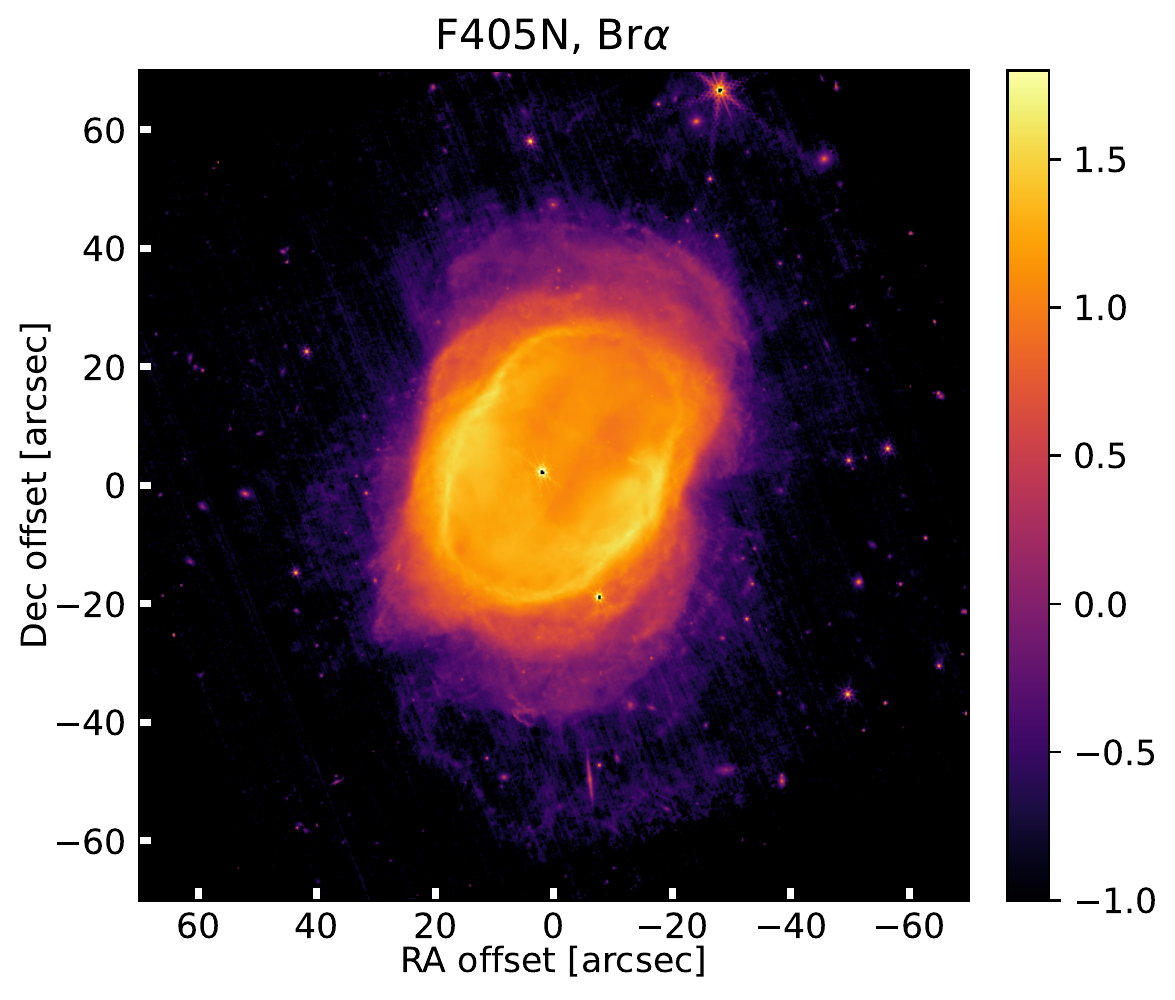}
	\includegraphics[width=3.9cm]{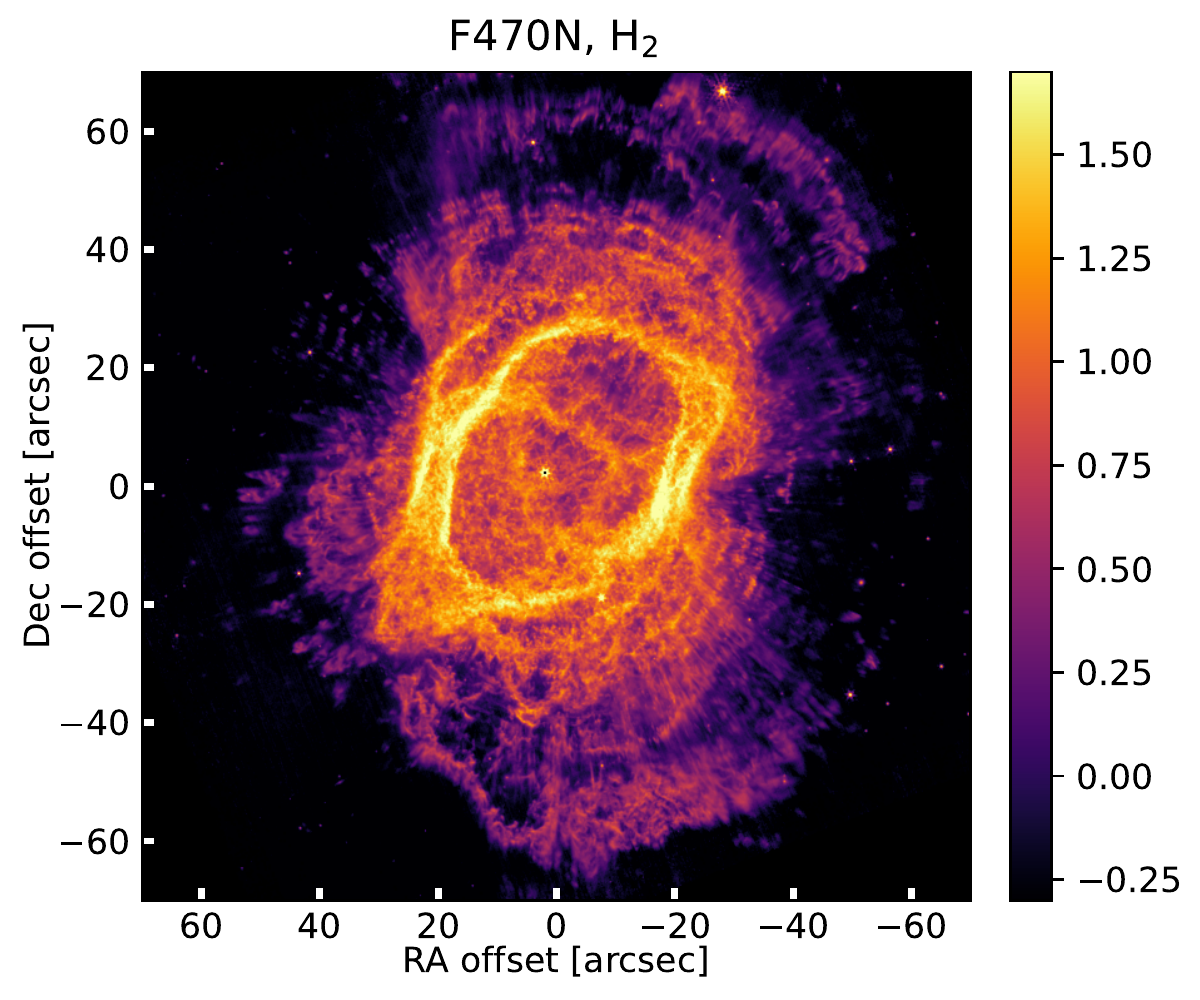}
	\includegraphics[width=3.9cm]{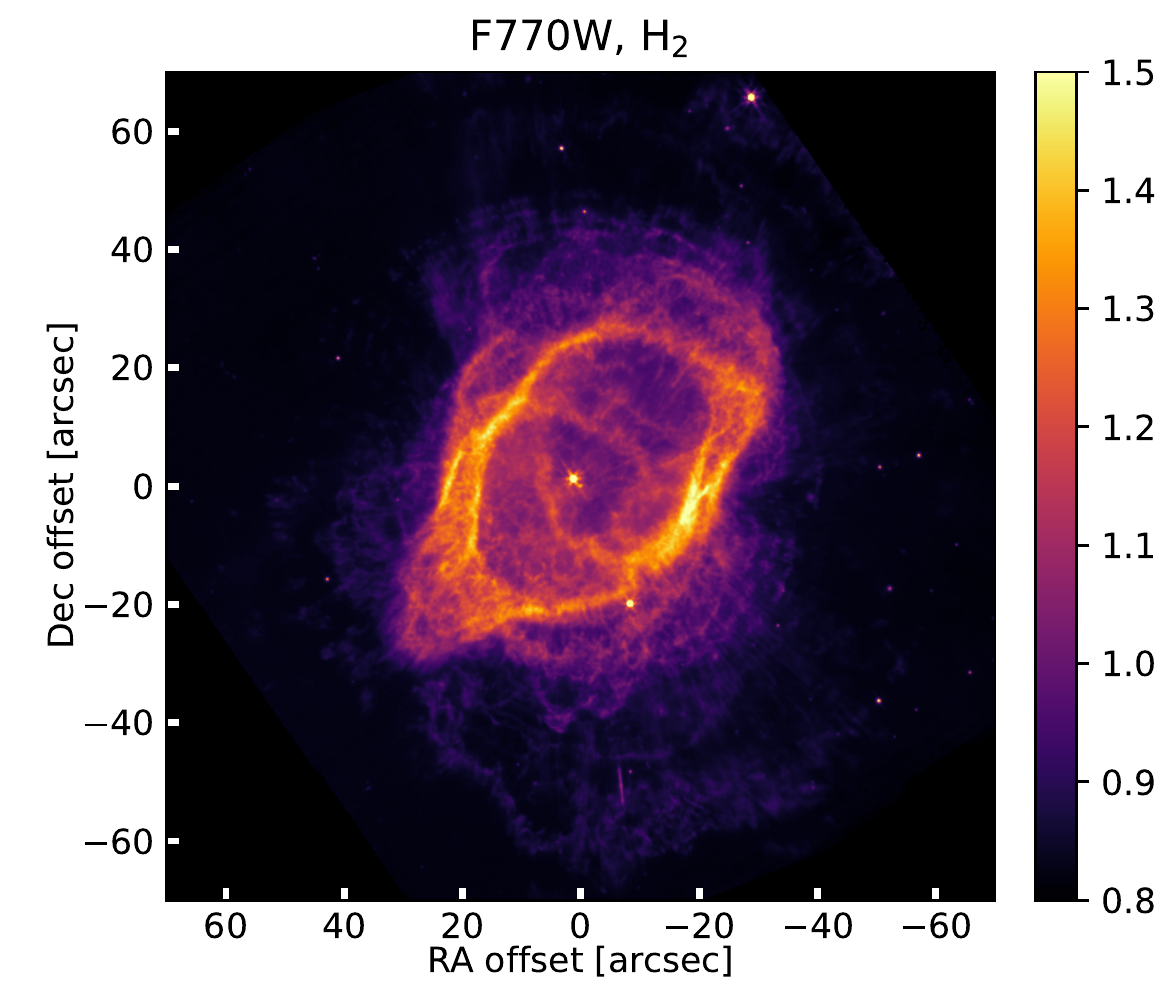}
	\includegraphics[width=3.9cm]{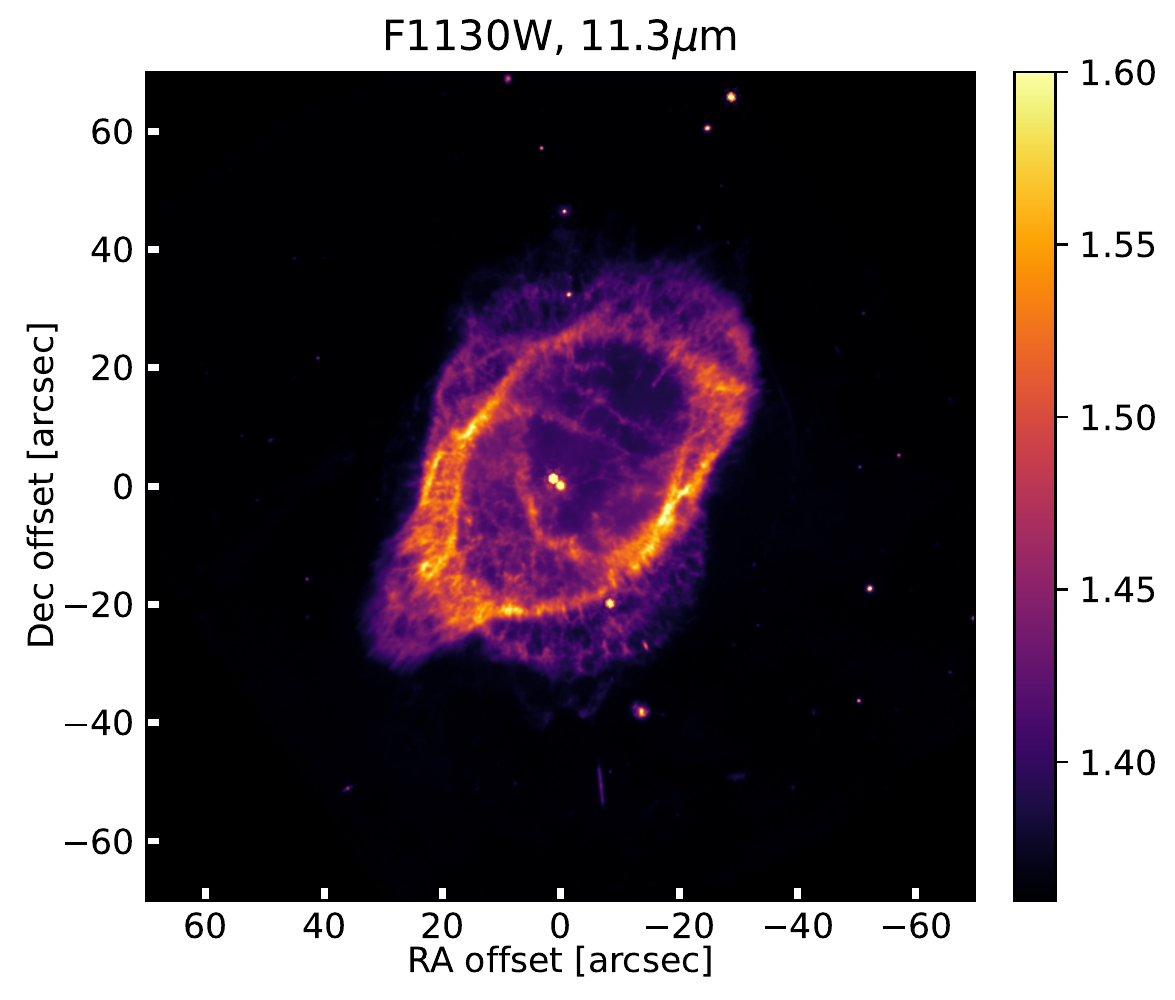}
	\includegraphics[width=3.9cm]{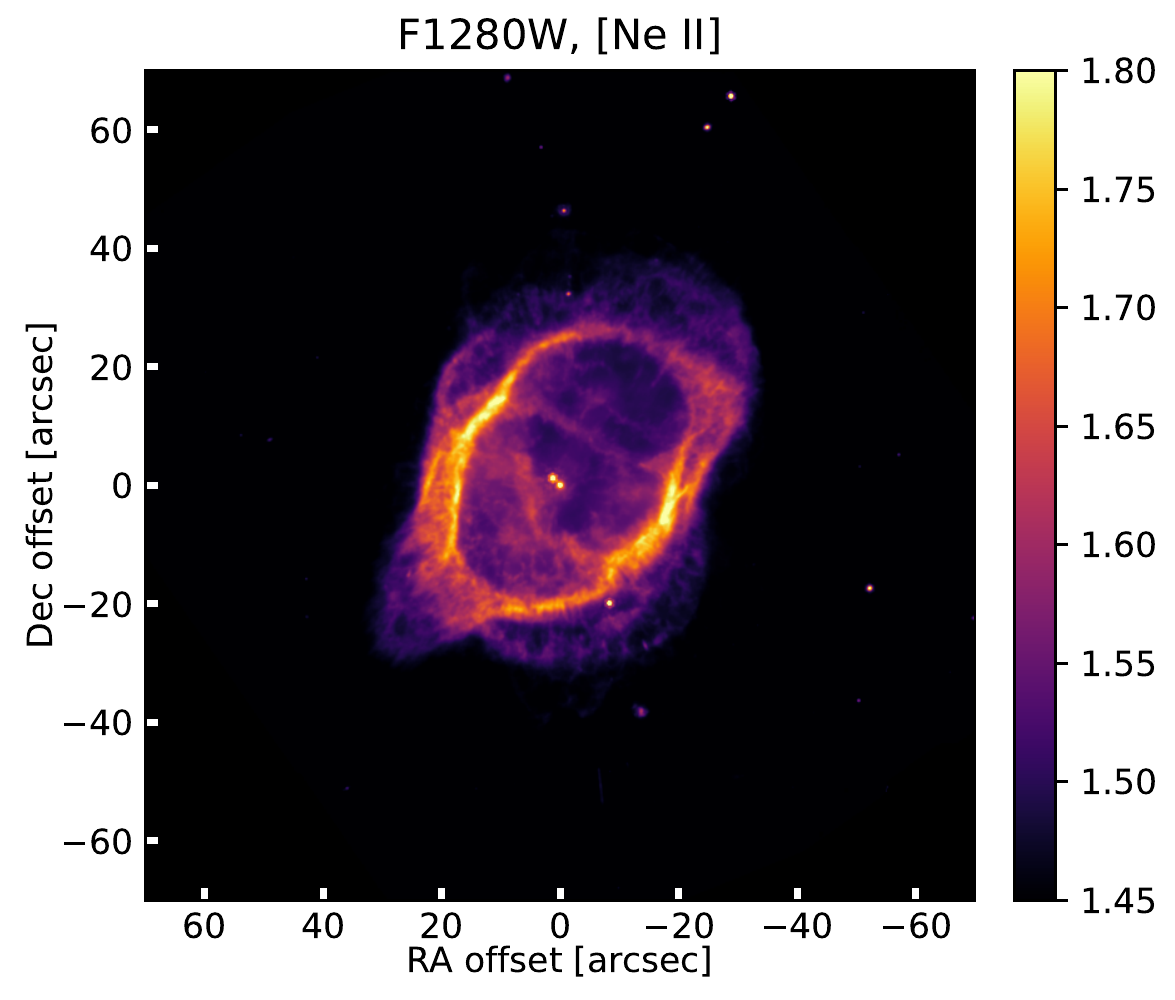}
	\begin{center}
	\includegraphics[width=3.9cm]{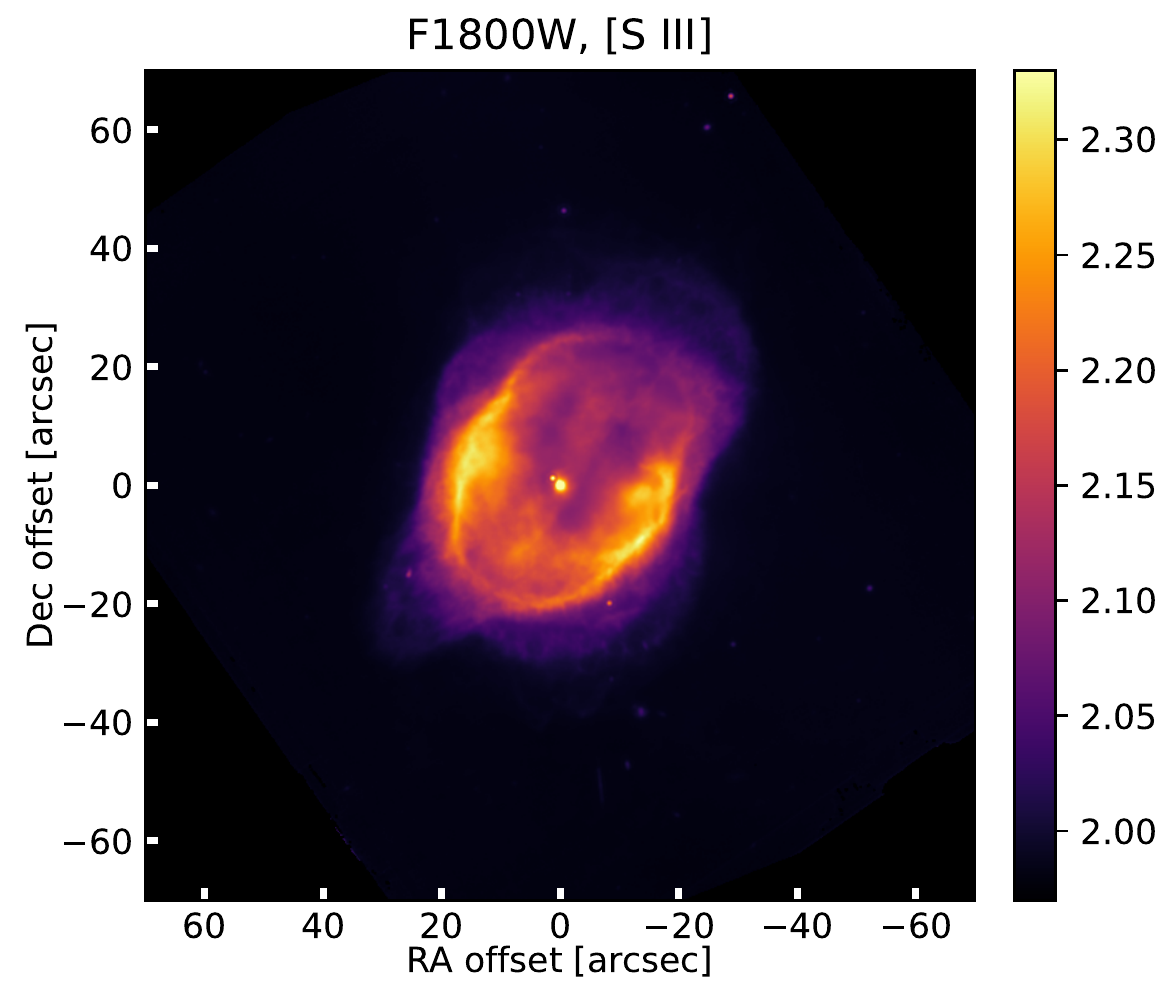}
	\end{center}
    \caption{JWST NIRCam and MIRI images of NGC~3132. North up and East is to the left. Colour bars indicate surface brightness in log(MJy~ster$^{-1}$).}
    \label{fig:postageStamps}
\end{figure*}

 \begin{figure}[htbp]
    \centering
    \includegraphics[width=9cm]{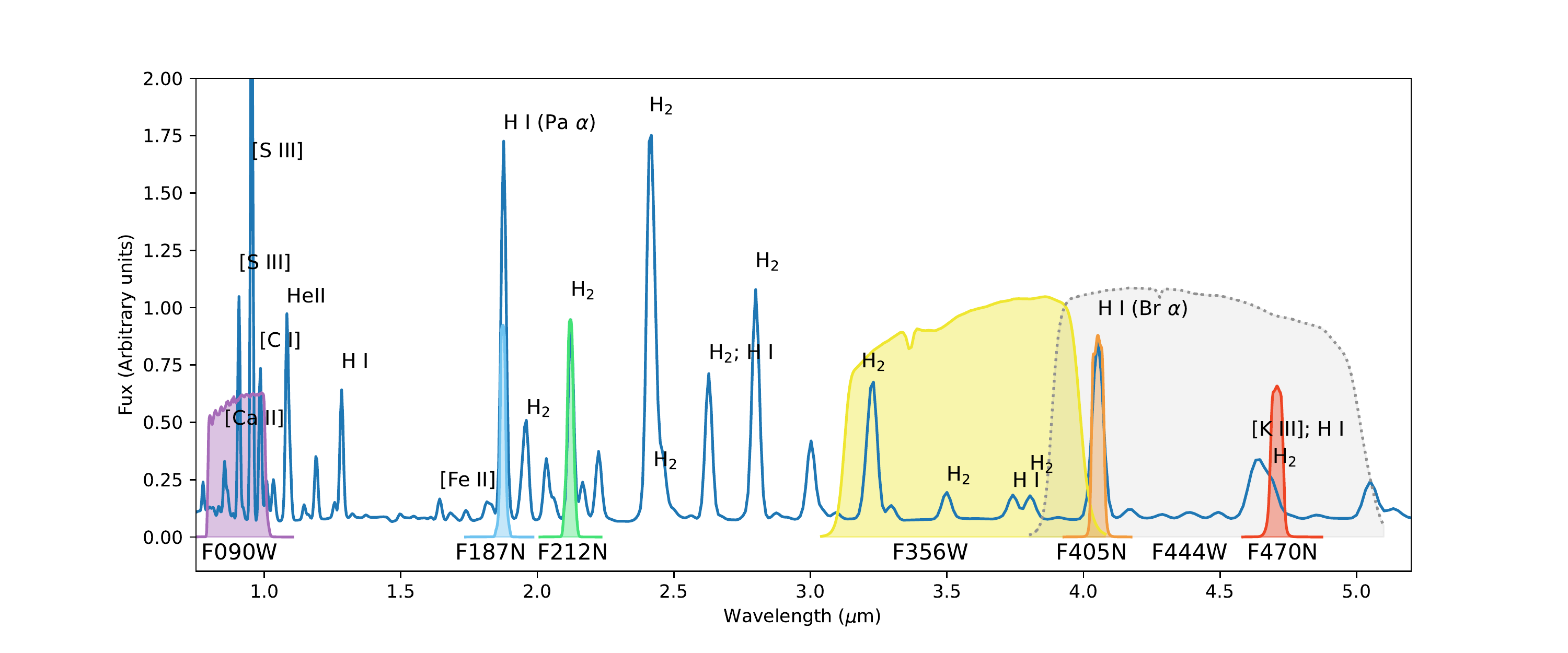}
    \includegraphics[width=9cm]{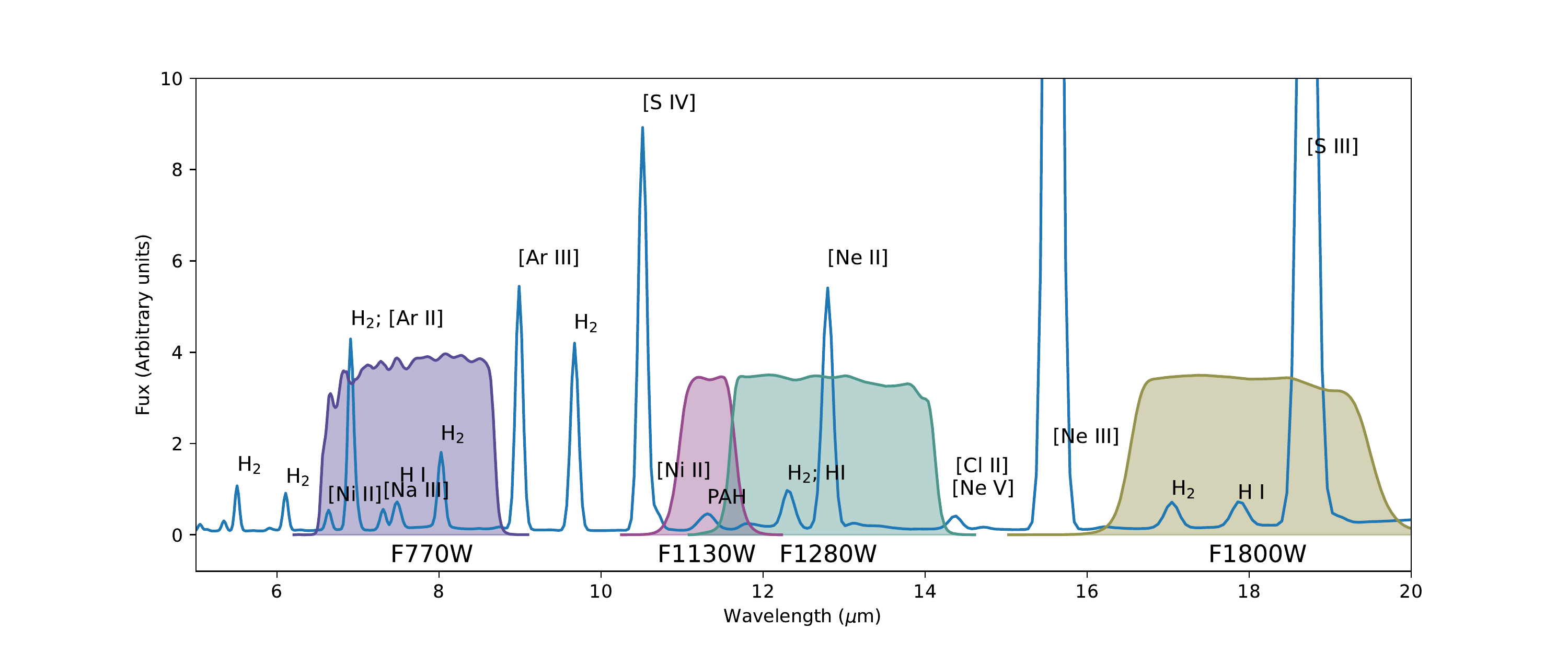}
    \caption{Simulated IR spectrum of NGC~3132, overlaid with the JWST filters to demonstrate bandpass contamination. Top panel: the simulated spectrum of NGC~3132 between 0.8 and 5.2~\mic\ (blue line) with, overlaid the JWST bandpasses (labelled coloured shapes). Bottom panel: the simulated spectrum of NGC~3132 between 5 and 20~\mic\ (blue line) with, overlaid the JWST bandpasses (labelled coloured shapes). }
    \label{fig:bandpass-contamination}
\end{figure}

\begin{figure}[htbp]
    \centering
    \includegraphics[width=9cm]{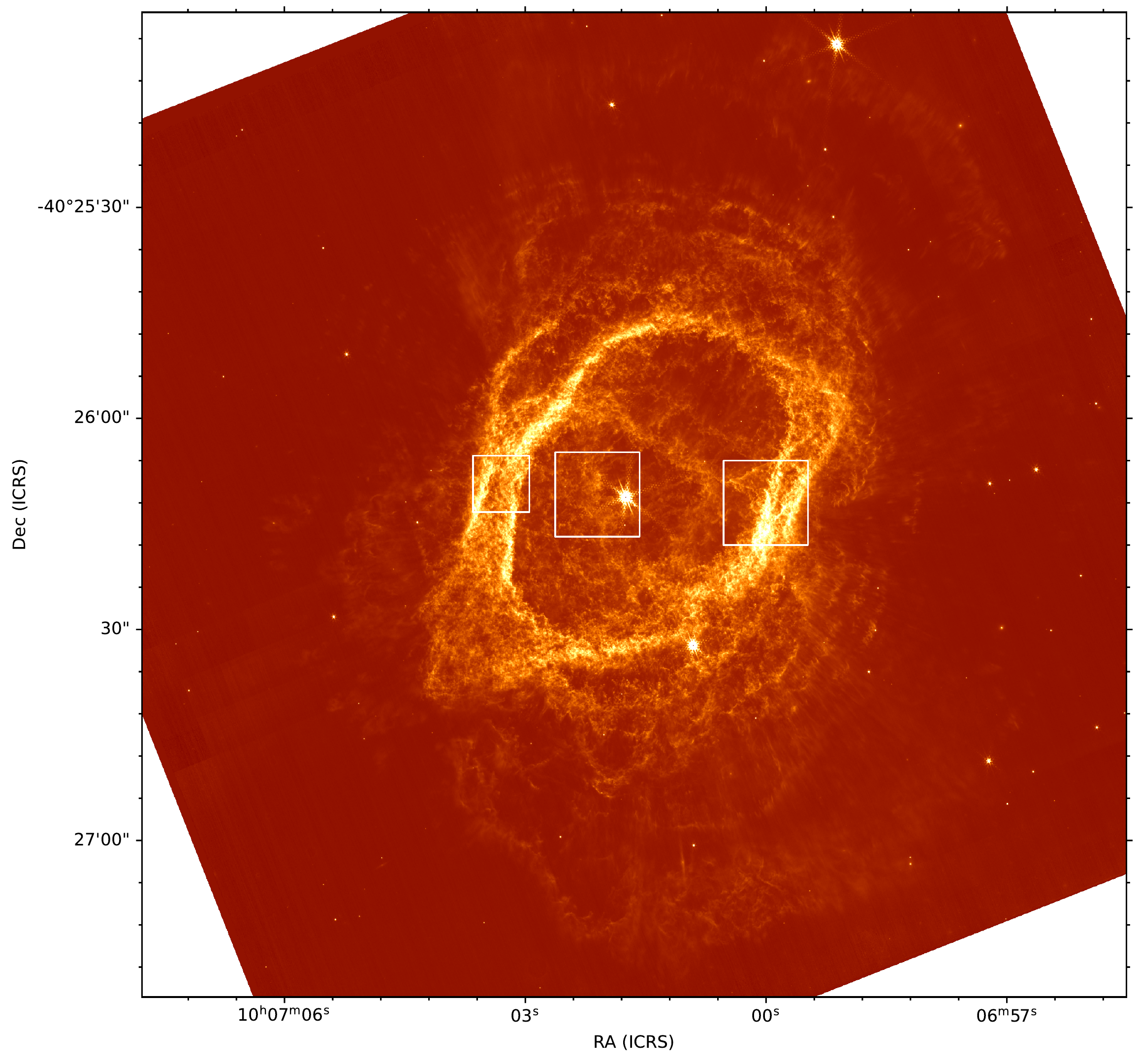}
    \caption{JWST/NIRCam F212N image of NGC~3132. White squares indicate positions and sizes of ``blowup'' regions highlighted in Figure~\ref{fig:MikakoRegions}.}
    \label{fig:MikakoH2}
\end{figure}

\begin{figure}[htbp]
    \centering
    \includegraphics[width=12cm]{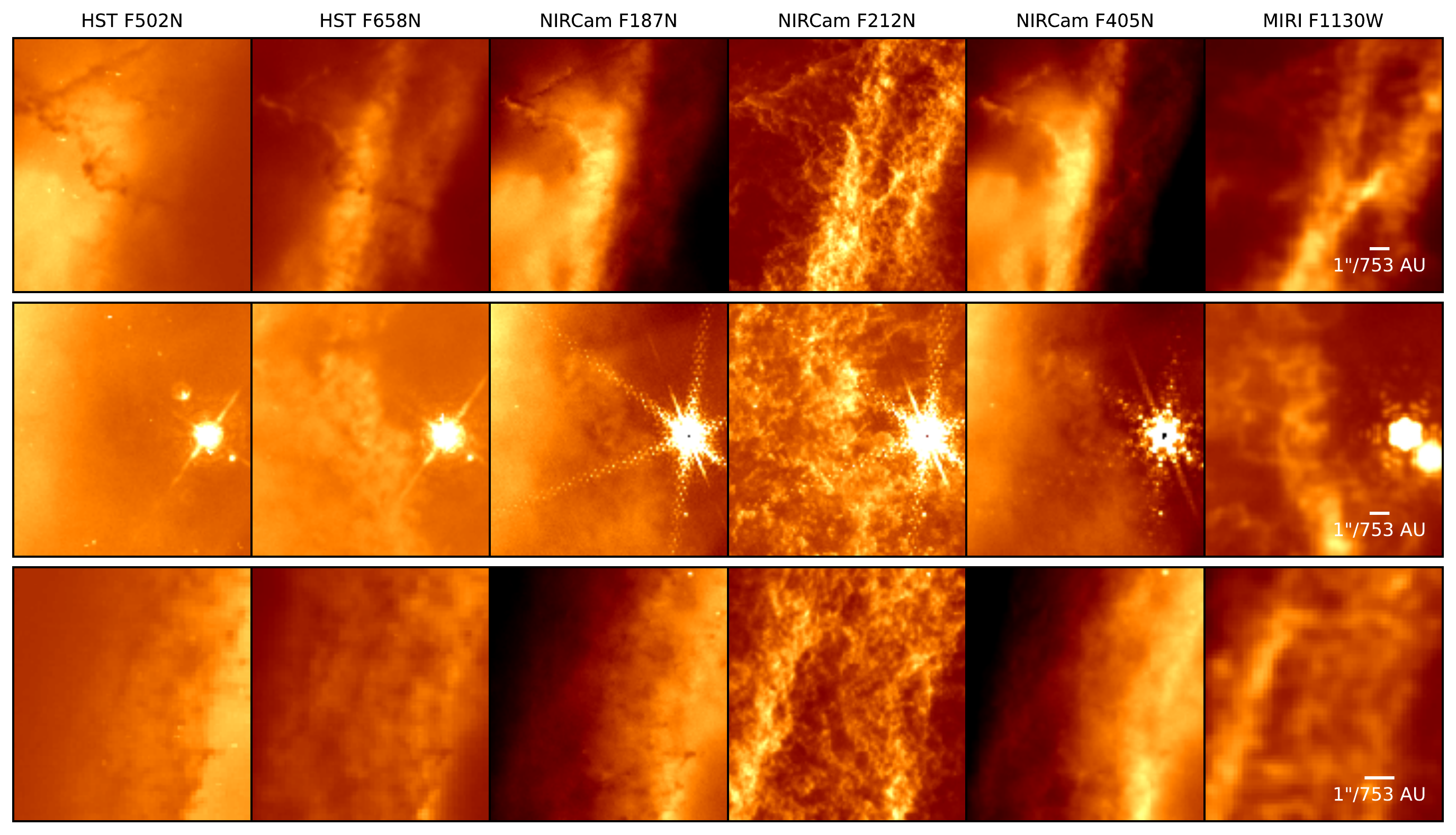}
    \caption{The enlarged images of knots in three representative regions, indicated by white boxes in Figure~\ref{fig:MikakoH2}: the West side of the ring (top row), near the centre of the ring (middle row) and the East side of the ring (bottom row). 
Filament structures stand out in the NIRCam F212N images. A few filaments are dusty, as clearly seen in the HST optical images (first two columns), but also in the NIRCam F187N and F405N images (third and fifth columns).
    \label{fig:MikakoRegions}}
\end{figure}

\subsection*{Central star magnitudes}
\label{sec:supplementary-central-star}
\begin{figure}[htbp]
    \centering
    \includegraphics{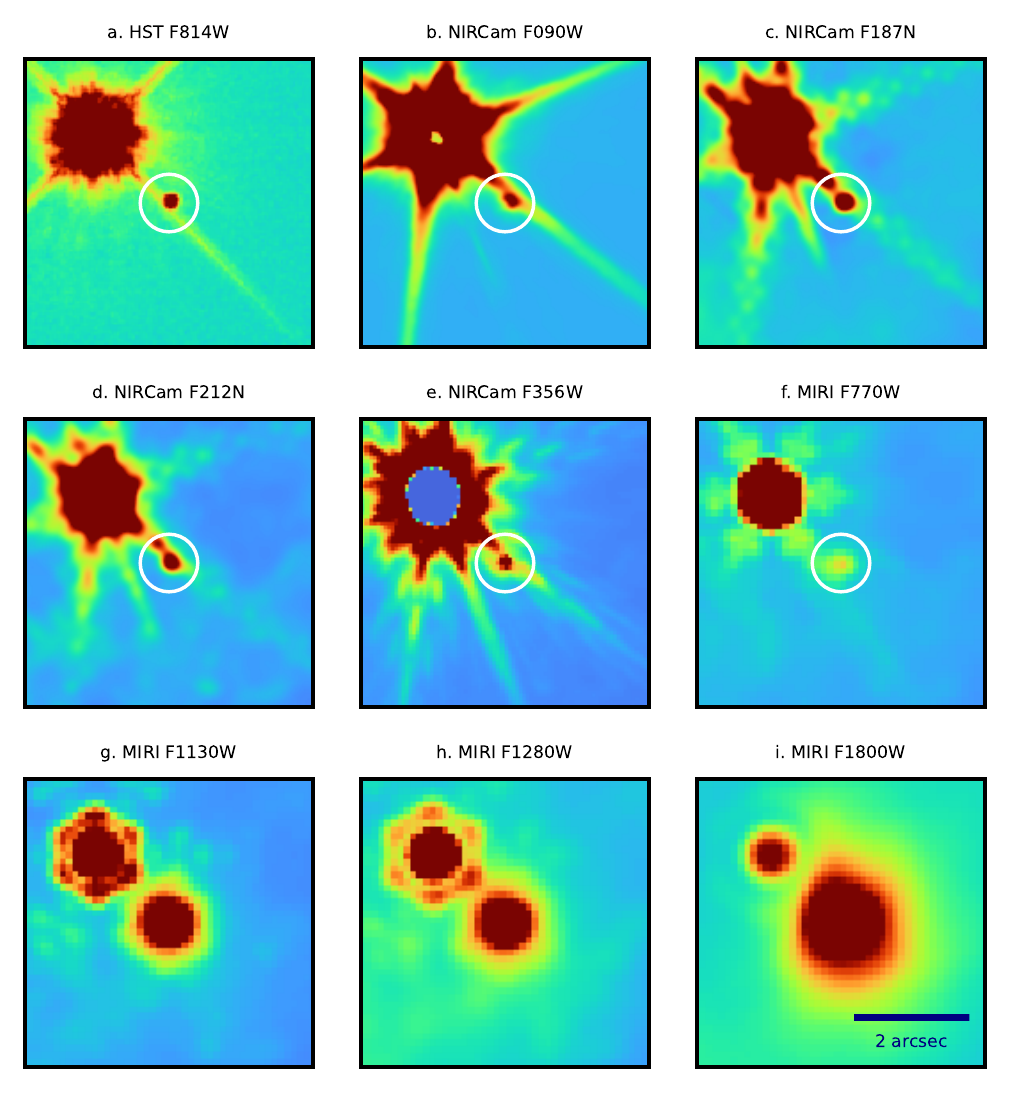}
    \caption{Sections of the JWST images zooming in on the central stars. NIRCam F090W, F187N and F212N images are deconvolved with simulated PSFs. HST, NIRCam F356W and MIRI images are not deconvolved. Note that the slight offset of the positions of the central star in the F090W and F187N images is due to imperfect deconvolution due to the nearby  saturated star.}
    \label{fig:central-star-postagestamps}
\end{figure}

JWST has detected the central visual binary from F090W to F1800W (Supplementary Figure~\ref{fig:central-star-postagestamps}).
At near-infrared wavelengths, the central star is faint and on the edge of the diffraction spikes from the nearby A-type star.
From 7.7\,$\mu$m longward, the central star is observed to increase in brightness, and at 18~$\mu$m it exceeds the flux from the  A-type star. 

We obtained optical magnitudes of the central star from the Hubble Source Catalogue version 3 \citep{2016AJ....151..134W}. The $F438W$, $F555W$ and $F814W$ Wide Field Camera 3 magnitudes in the AB system were converted to Jy. 
The calibrated 2D resampled NIRCam and MIRI observations available as i2d pipeline products  were used to perform aperture photometry of the central star using the \textsc{photutils} package \citep{photutils}. A circular aperture was centered on the central star with radii corresponding to the 80\% encircled energy radius tabulated in the relevant aperture correction tables, sufficient to include the majority of the central star flux. The tables JWST\_nircam\_apcorr\_0004 and JWST\_miri\_apcorr\_0008 were sourced from the JWST Calibration Reference Data System. Aperture photometry of the central source was performed with the error extension of the image included. Three circular sky apertures of the same radius were selected nearby the central star to best sample the challenging background. The background includes artefacts from the diffraction pattern of the A-type star nearby (more prominent in the NIRCam images) and the structured nebular background from NGC~3132. In the $F090W$, $F212N$, $F405N$ and $F470N$ filters the dominant background and/or intrinsic faintness of the central star precluded any meaningful fluxes from being measured.

The sky background was estimated as the average of the median counts in each sky aperture, where the median was calculated using sigma clipping with $\sigma=3$. The sky background was scaled to the aperture area $A$ of the central star aperture before it was subtracted from the aperture sum. The flux of the central star was calculated as the sky subtracted aperture sum scaled by the MJy/sr to $\mu$Jy conversion factor and the aperture correction sourced from the aperture correction tables. The uncertainty in the flux was estimated as $\sqrt{\sigma_p + 2\sigma_\mathrm{sky}}$ where $\sigma_p$ is the \textsc{photutils} aperture\_sum\_err and $\sigma_\mathrm{sky}$ is $A\sigma_b^2$, where $\sigma_b$ is the average of the standard deviation of counts in each sky aperture. However, due to the complex sky background, these uncertainties are likely underestimates of the true uncertainty, which we estimate to be 5--10\% of the flux.

Supplementary Table \ref{tab:fluxes} gives the measured and dereddened fluxes using $A_V=3.1E(B-V)$, where $E(B-V)=0.09$ mag \citep{MonrealIbero20}, and filter extinction ratios are from the Spanish Virtual Observatory Filter Profile Service \citep{svo1,svo2}.

In Supplementary Figure~\ref{fig:hrA} we present the PARSEC isochrones with the derived location of the A2V star.
\begin{figure}[htbp]
\centering
	\includegraphics[width=9cm]{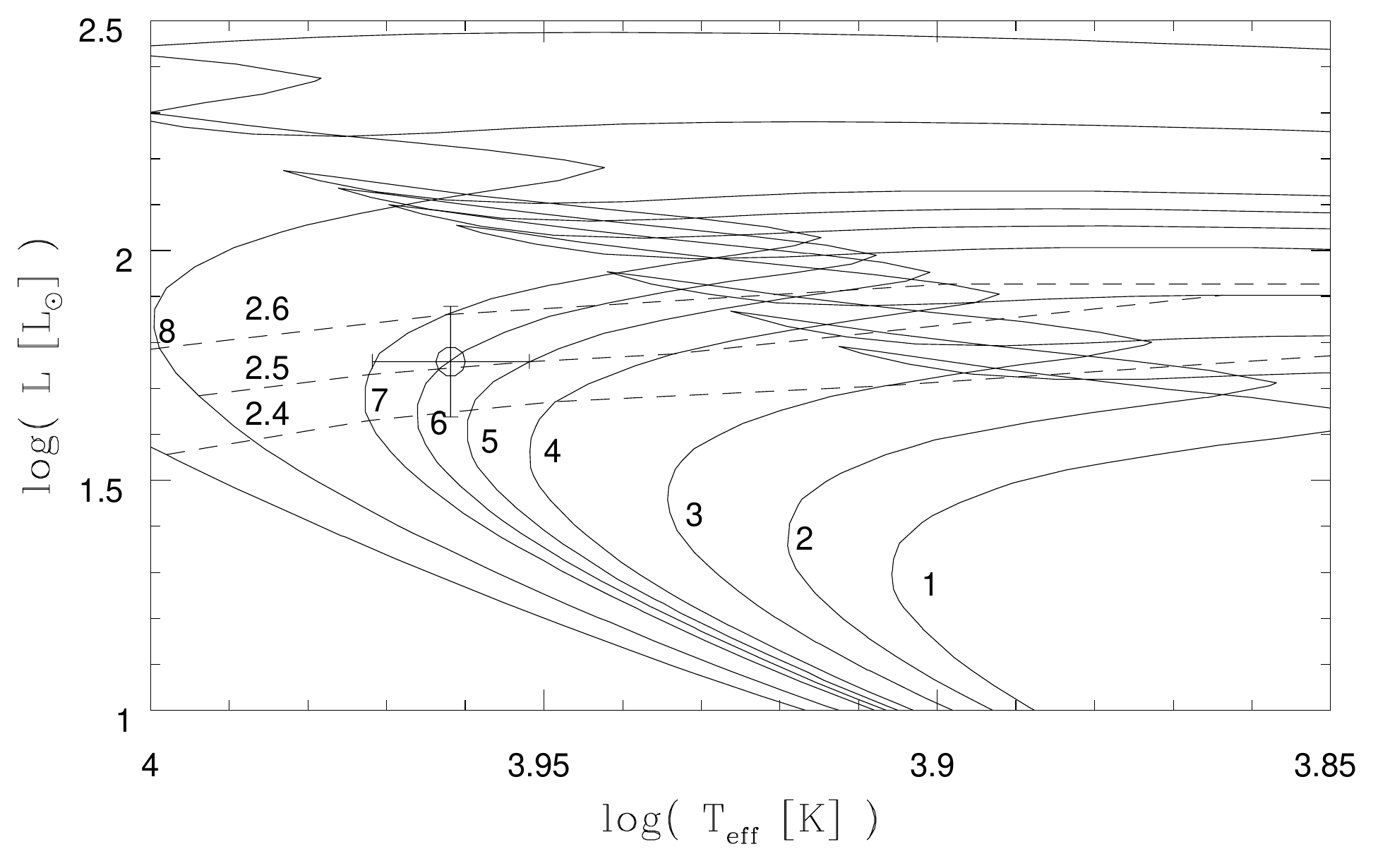}
    \caption{PARSEC \cite{Marigo2017} isochrones with the location of the A2~V star. The isochrones are labelled in order of descending age  : 9, 8, 7, 6, 5.6, 5.3, 5, 4 in units of $10^8$\,yr. The dashed lines connect points of constant mass, labelled in solar units. The error bar on the horizzontal axis is the temperature uncertainty given by {\it Gaia}, $\pm$200~K. The error bar on the vertical axis is based on a conservative 0.25~mag error on the absolute magnitude.}
    \label{fig:hrA}
\end{figure}

\begin{table}[htbp]
    \centering
    \begin{tabular}{lrccc}
    \hline
        Filter & $F_\nu$ ($\mu$Jy) & Formal  & $F_{\nu,0}$ ($\mu$Jy) & Formal \\
         &  & error$^*$  &  & error$^*$\\
        \hline
WFPC3/F438W & 1620.32 & 10.21 & 2274.64 & 14.3 \\
WFPC3/F555W & 1188.50 & 7.33 & 1556.61 & 9.6\\
WFPC3/F814W & 560.27 & 3.29 & 654.18 & 3.8\\
NIRCam/F187N & 130.43 & 4.65 & 135.77 & 4.84\\
NIRCam/F356W & 50.78 & 5.31 & 51.67 & 5.41\\
MIRI/F770W & 285.68 & 13.47 & 288.02 & 13.58\\
MIRI/F1130W & 1445.32 & 2.46 & 1462.85 & 2.49\\
MIRI/F1280W & 1362.75 & 5.85 & 1373.09 & 5.90\\
MIRI/F1800W & 11344.60 & 20.93 & 11425.35 & 21.08\\
\hline
\multicolumn{5}{l}{$^*$The actual error on this photometric measurements is likely}\\
\multicolumn{5}{l}{$^*$ closer to 5-10\% of the flux values.}\\
    \end{tabular}
    \caption{Measured ($F_\nu$) and dereddened (F$_{\nu,0}$) fluxes of the central star. To convert to magnitudes: $mag (AB) = -2.5 * log10(F_\nu\ [{\rm in\ Jansky}]) + 8.90$}
    \label{tab:fluxes}
\end{table}

\subsection*{Extended structure of the central star }
\label{sec:supplementary-central-star_F1800W}

\begin{figure}
\centering
\includegraphics[width=9cm, trim={0.7cm 0.2cm 1.5cm 1.2cm},clip]{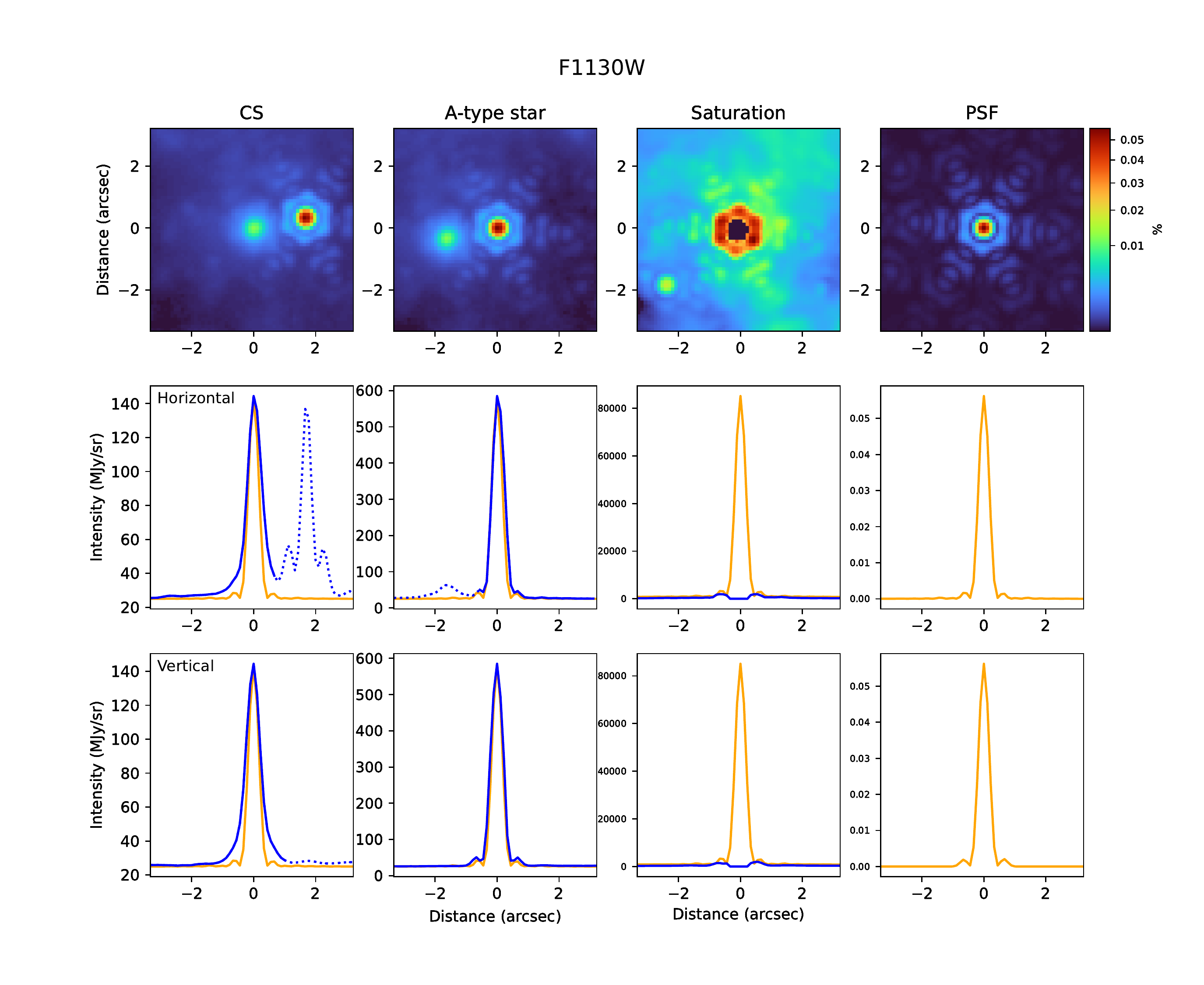}
\includegraphics[width=9cm, trim={0.7cm 0.2cm 1.5cm 1.2cm},clip]{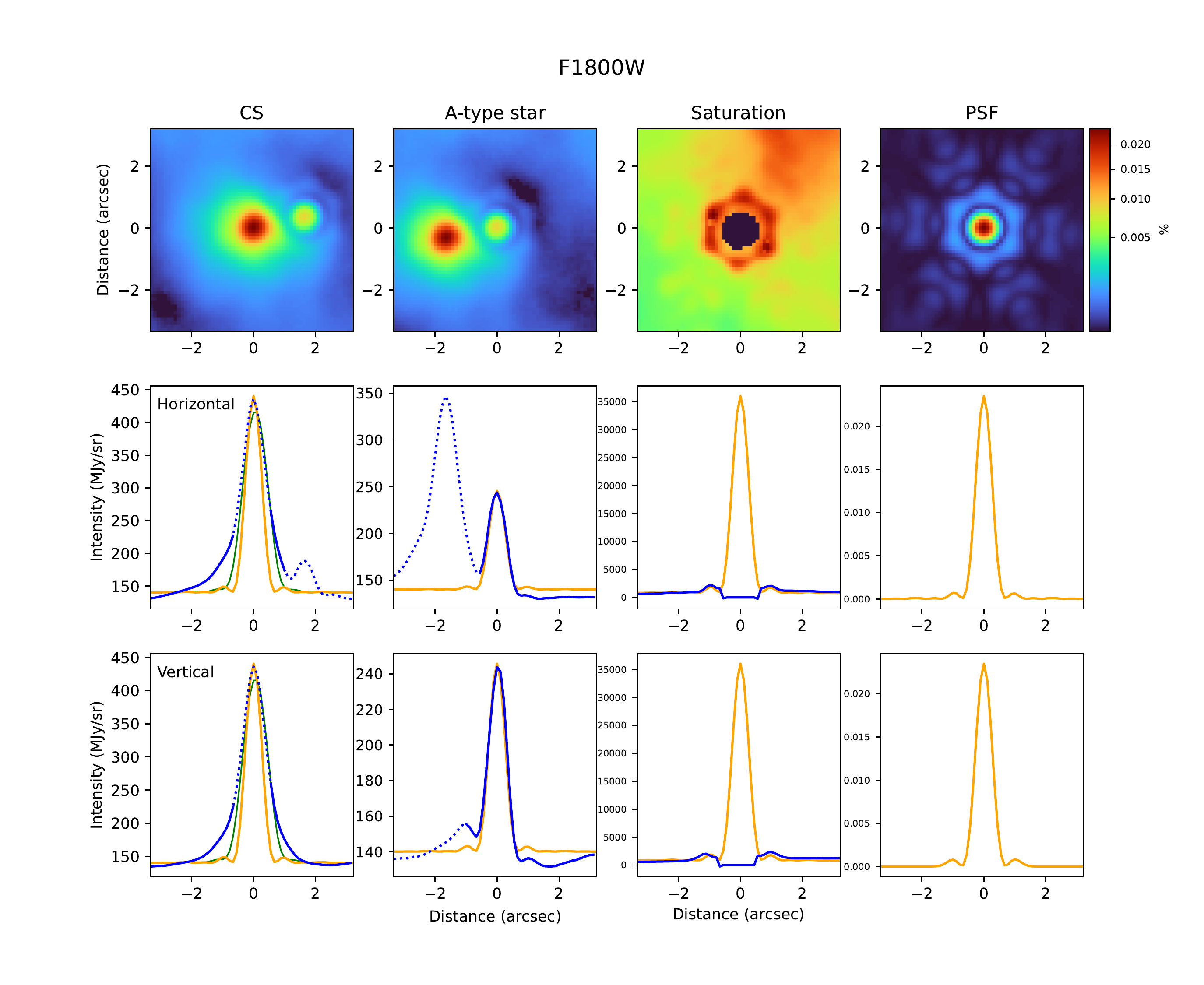}
\caption{Demonstration of the extended structure of the central star in the MIRI F1130W (top panel) and F1800W (bottom panel) band images. From the left to right column, the central star, the A-type star, an example of a saturated star, and the simulated PSF. The top row displays the images, and the second and third row show radial profiles, which are sliced in the horizontal and vertical directions across the peak (blue lines). Dotted lines indicate the data points strongly affected by other factors, such as bright neighbouring stars or detector saturation for the F1800W image. In the radial profiles, the PSF is also plotted as an orange line.
The central star is clearly extended more than the PSF. On the central star radial profiles of the F1800W image, the green line demonstrates the radial profile of a 0.44\,arcsec-radius, flat-intensity ‘disk’. \label{fig:NGC3132_CS_F1800W}}
\end{figure}

Supplementary Figure\,\ref{fig:NGC3132_CS_F1800W} demonstrates the extended structure of the central star in the MIRI F1130W and F1800W band images. From the left to right column, we see the central, the A-type star, an example of a saturated star, and the simulated PSF. The images are oriented such that North is 145\,degrees clockwise. An example image of a saturated star is from the Tarantula Nebula region, which was observed as a part of the first image program (PID 2729). The coordinate of this saturated star is RA=05h38m33.61s and Dec=$-$69d04m50.5s. The MIRI PSF is simulated  based on Webb PSF software version 1.1.0 (https://jwst-docs.stsci.edu/jwst-mid-infrared-instrument/miri-performance/miri-point-spread-functions). The top row shows the images, and the second  and third rows show radial profiles, which are sliced in horizontal and vertical directions across the peak (blue lines). Dotted lines indicate the data points  that are strongly affected by other factors, such as bright neighbouring stars or detector saturation. The central star is mildly saturated in the F1800W image, with data quality flags of about 5—8 pixels across the peak, so that the data within the 10 pixels from the peak of the central star are plotted as dotted lines. On the central star radial profiles of F1800W, green lines demonstrate the simulated radial profile with 0.44\,arcsec radius flat-intensity ‘disk’.  This shows that the central star is extended at a scale of $\gtrsim$0.8\,arcsec. In the radial profiles, the PSF is also plotted as an orange line, which has a FWHM of 0.58\,arcsec at F1800W. 

Supplementary Figure\,\ref{fig:NGC3132_CS_F1800W} demonstrates that the central star is clearly extended more than the PSF. On the central star radial profiles, green lines demonstrate the radial profile with a 0.44\,arcsec radius, flat-intensity ‘disk’. 
In reality, the intensity gradually decreases radially, rather than a flat intensity with a cliff edge, and the tailing of this gradual decrease continues beyond 0.4\,arcsec.

\subsection*{Morpho-kinematic modelling}
\label{sec:supplementary-morpho-kinematic-modelling}

In Figure~3 we have presented the 3D morpho-kinematic reconstruction of the   ionised region of NGC~3132 using the interactive morpho-kinematic modelling software \textit{Shape} \citep{Steffen2011}. In addition to the new images from JWST, spectroscopic reference data for the reconstruction are position-velocity diagrams from the San Pedro M{\'a}rtir Kinematic Catalogue of Galactic Planetary Nebulae (\cite{lopez2012san,Hajian2007}).

An assumption is made on the current velocity field in order to map the Doppler-shift to a 3D position of that image element. We assume an overall homologous velocity field \citep{Zijlstra2001}, except locally for some protrusions (see below). The velocity field is 1~km~s$^{-1}$~arcsec$^{-1}$. This value ensures that the cross-section of the main shell is approximately circular. We estimate the uncertainty to be of the order of 30\%. Note that the stretching of the structures along the line of sight is proportional to this value. In other words, the shape of the nebula along the line of sight is linearly related to the component of the velocity vector. This model of the inner ionised cavity supersedes or rather, completes, the barrel or ``diabolo'' model \cite{Monteiro2000} in view of more sensitive spectroscopy that allowed us to detect the faint, and fast, closed ends of the ellipsoid along its major axis, which is close to the line of sight.

In Supplementary Figure~\ref{fig:de-project_shape} we show the slit positions and resulting position-velocity diagram (top row: observed, bottom row: simulated) reconstruction with \textit{Shape}. A 3D fly through the 3D volume can be found at this \href{www.ilumbra.com/public/science/ilumbra_NGC3132_NII_reconstruction.mp4}{link}. The position-velocity diagrams in Supplementary Figure~\ref{fig:de-project_shape} are effectively 2D renditions of the spectral line shape as we move along the slit. The slit positions are indicated. 

In the final step of our morpho-kinematic model (Figure~3),
we place two complete shells of non-uniform, filamentary material around the central star to match the observed size of the \hh\ halo. The inner shell ranges from 33 to 45~arcsec from the central star, and the second ranges from 60 to 70~arcsec. These are illuminated through a partially opaque ellipsoid (approximately corresponding with the ellipsoid containing the ionised nebula) that has reduced opacity around the poles and has an overall porous opacity. 

At this point we are only interested in  testing the possibility that the opacity of the walls of the central cavity within which the exciting central star resides, could be the cause of the overall emissivity distribution and features. The central star is made to radiate as a blackbody. We have then used a simple proxy for the various radiative processes that are at work here, i.e., isotropic scattering on dust.  Since we expect dust to dominate the transport of the exciting radiation, this is a reasonable first test, with the details being irrelevant for this simple geometric simulation. A spherical density modulation was also imposed with {\it ad hoc} spacing responding to the following modulation: $\rho / \rho_0 = 0.3+\sin(0.8\ r/{\rm arcsec})^{10}\ \sin(1.4\ r/{\rm arcsec})^2$. This generates the arch pattern.

This type of painstaking morpho-kinematic modelling is critically dependent on images at different wavelengths as well as spatially-resolved spectroscopy. On the basis of this type of data driven 3D model, we can now understand structures first revealed by JWST such as the \hh\ halo and the dusty central stars.

\begin{figure}[htbp]
\centering
\includegraphics[width=7cm]{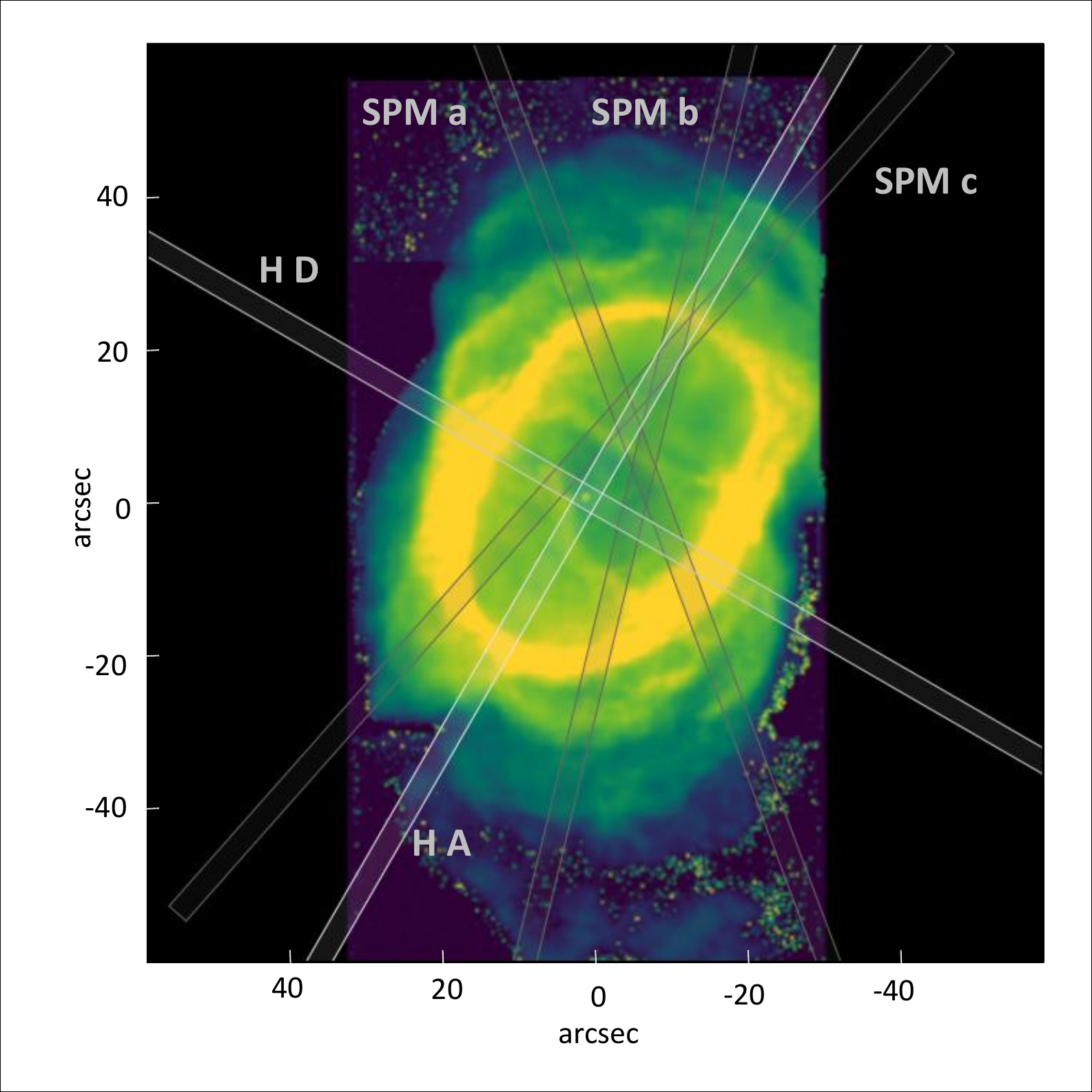}
\includegraphics[width=7cm]{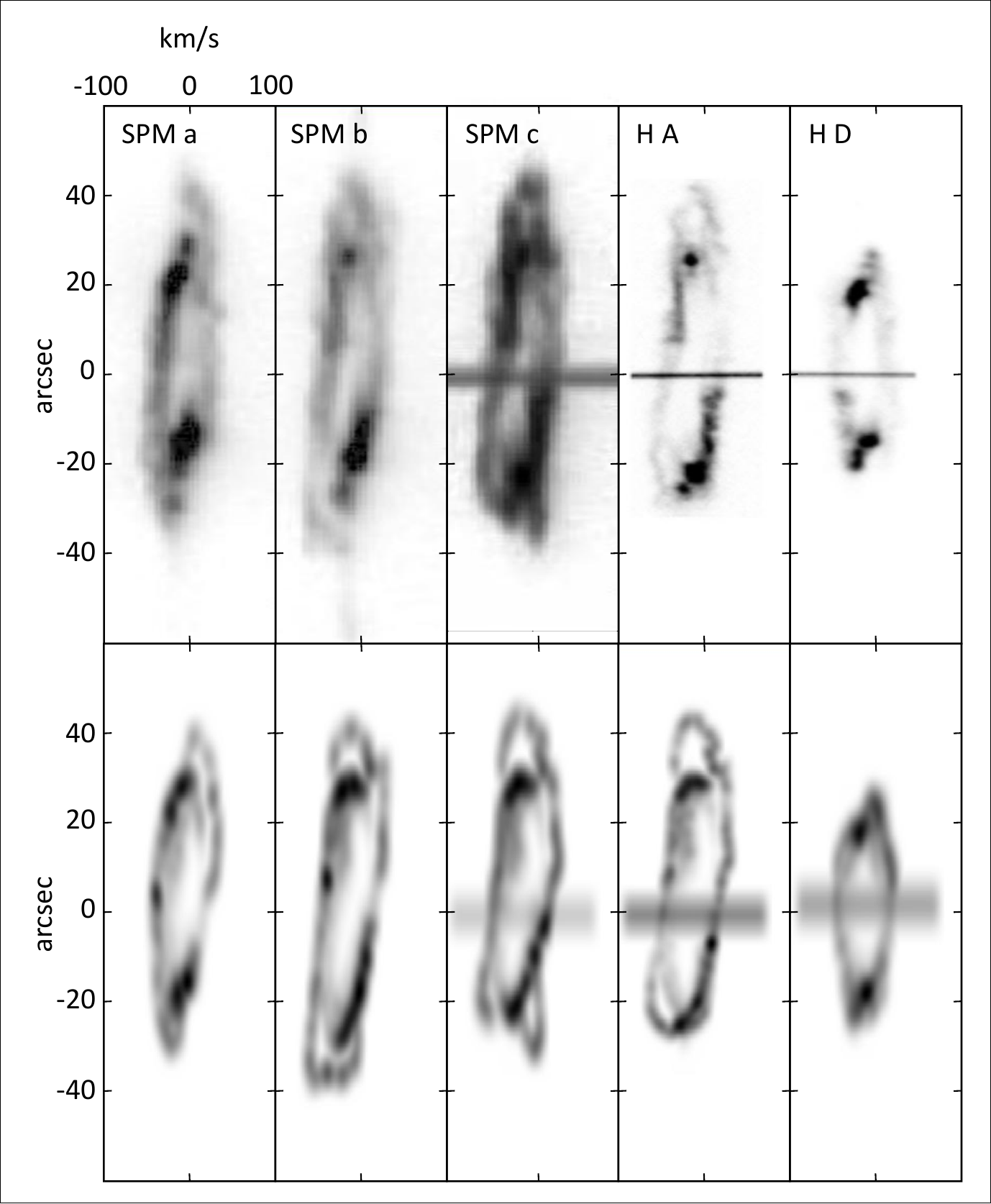}
  \caption{Position velocity data and model to achieve the morpho-kinematic model of PN NGC~3132. Top figure: a MUSE [\nii] image of NGC~3132 with the slit positions marked. The three slits from \cite{lopez2012san} are marked in grey and the two slits from \cite{Hajian2007} are in white. Bottom figure, top row: observed position-velocity diagrams for the [\nii] nebular line along the different slits. Bottom figure, bottom row: corresponding modelled position-velocity diagrams using {\it Shape}. The heavy horizontal lines in the last three columns are the continuum emission from the A2~V star.} 
  \label{fig:de-project_shape}
\end{figure}

\end{appendices}


